\documentclass[acmsmall,authorversion]{acmart}

\usepackage{booktabs} 
\usepackage{multirow}
\usepackage{graphicx}
\usepackage{subcaption}

\newcommand{\faUserO}{\includegraphics[keepaspectratio=true,width=3mm]{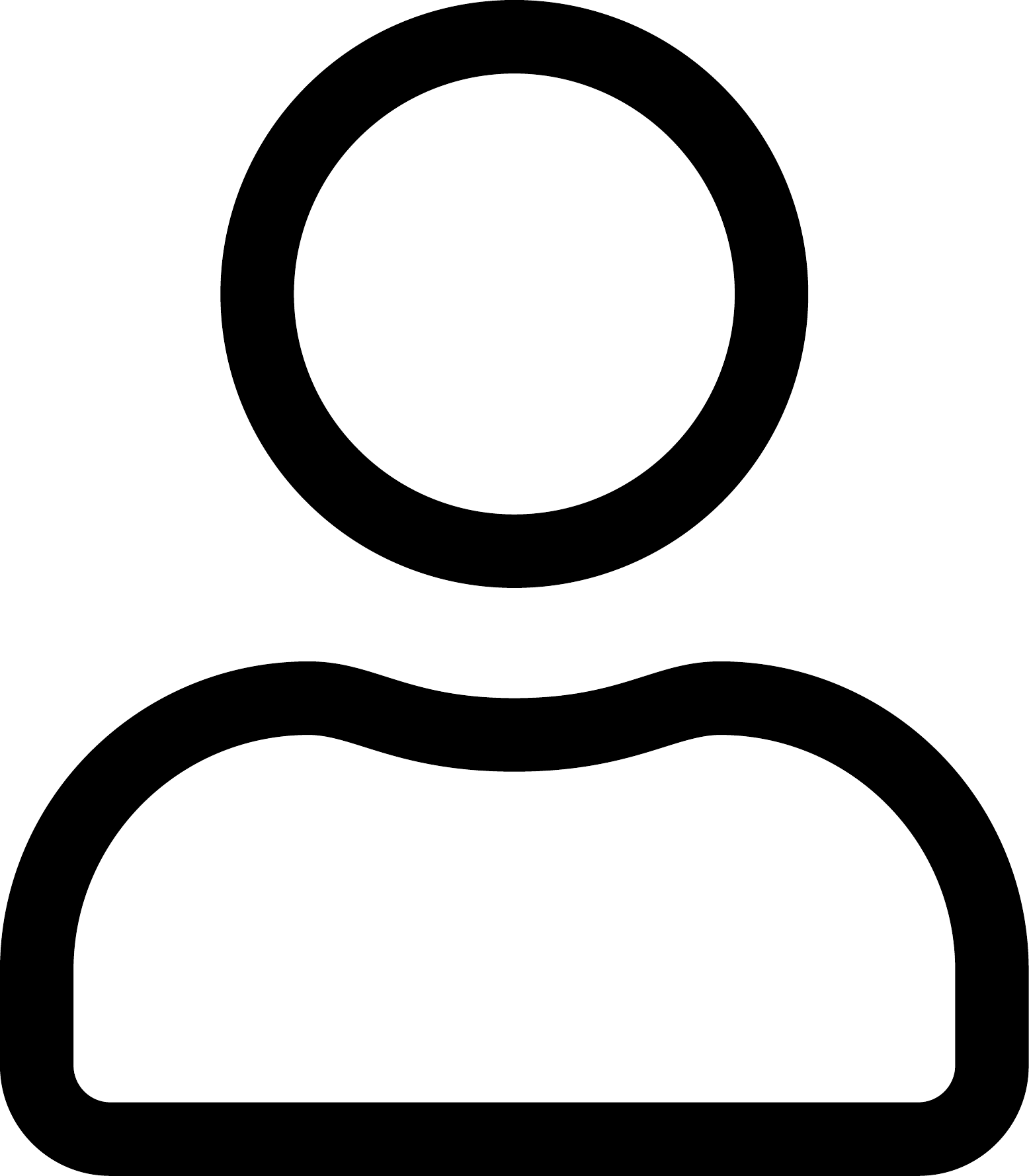}}
\newcommand{\faUserOSmall}{\includegraphics[keepaspectratio=true,width=2.1mm]{pic/fa-user.pdf}}
\newcommand{\faDesktopS}{\includegraphics[keepaspectratio=true,width=3.5mm]{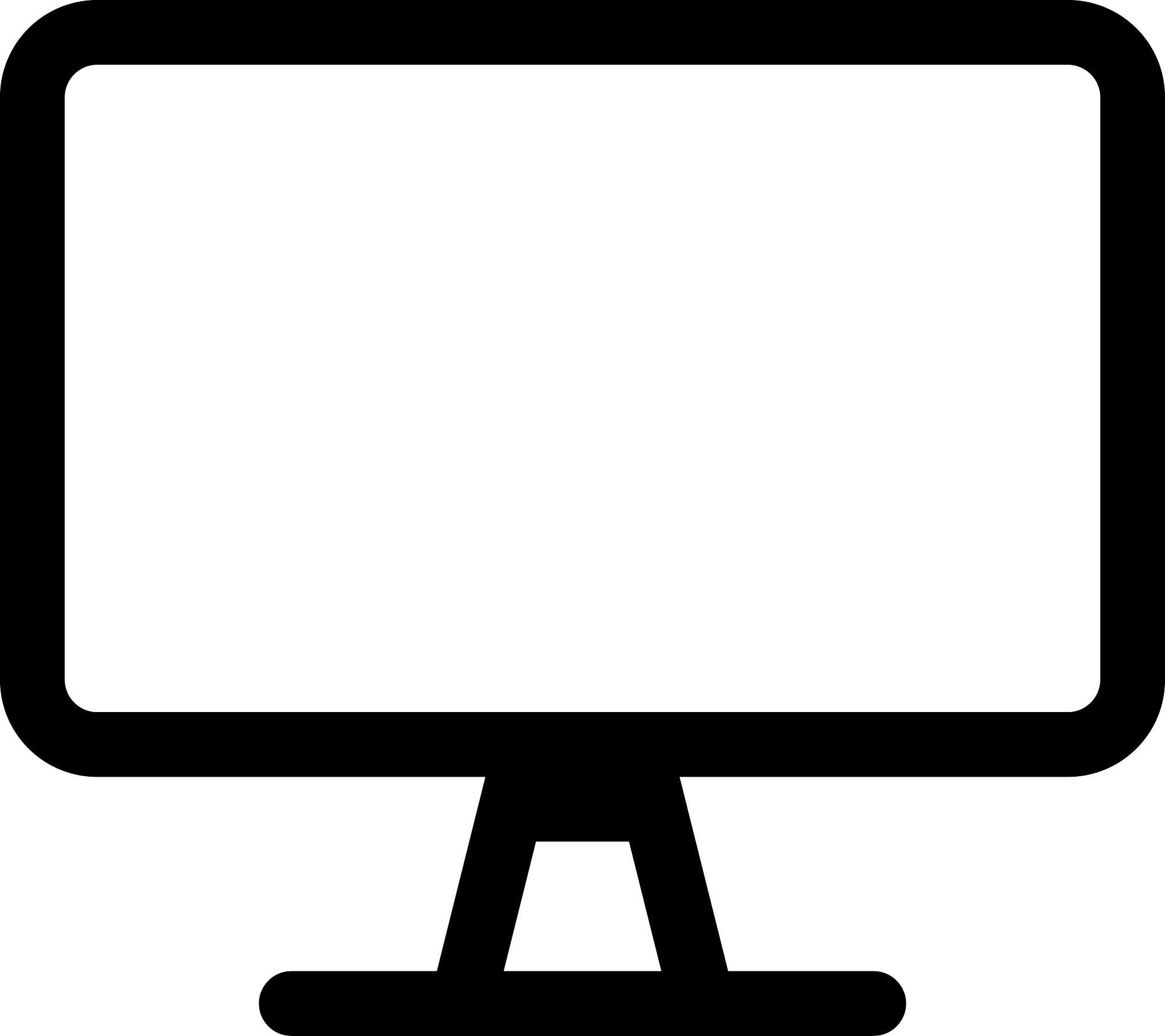}}
\newcommand{\faDesktopSmall}{\includegraphics[keepaspectratio=true,width=2.4mm]{pic/fa-desktop.pdf}}

\usepackage[normalem]{ulem}

\newsavebox{\imagebox}

\usepackage{array}
\newcolumntype{L}[1]{>{\raggedright\let\newline\\\arraybackslash\hspace{0pt}}m{#1}}
\newcolumntype{C}[1]{>{\centering\let\newline\\\arraybackslash\hspace{0pt}}m{#1}}
\newcolumntype{R}[1]{>{\raggedleft\let\newline\\\arraybackslash\hspace{0pt}}m{#1}}
\usepackage{tabularx}
\usepackage{xcolor,colortbl}

\setcopyright{acmlicensed}
\acmJournal{PACMHCI}
\acmYear{2019} \acmVolume{3} \acmNumber{CSCW} \acmArticle{195} \acmMonth{11} \acmPrice{15.00}\acmDOI{10.1145/3359297}

\received{April 2019}
\received[revised]{June 2019}
\received[accepted]{August 2019} 

\begin{document}
\title[Discovering The Sweet Spot of Human---Computer Configurations]{Discovering the Sweet Spot of Human---Computer Configurations: A Case Study in Information Extraction}
%

\author{Maximilian Mackeprang}\authornote{Both authors contributed equally to the paper.}
\email{maximilian.mackeprang@fu-berlin.de}
\affiliation{%
  \institution{Human-Centered Computing, Freie
Universit\"at Berlin}
  \streetaddress{14195 Berlin, Germany}
}

\author{Claudia M\"uller-Birn}\authornotemark[1]
\email{clmb@inf.fu-berlin.de}
\affiliation{%
  \institution{Human-Centered Computing, Freie
Universit\"at Berlin}
}

\author{Maximilian Stauss}
\email{max.stauss@fu-berlin.de}
\affiliation{%
  \institution{Human-Centered Computing, Freie
Universit\"at Berlin}
}


\begin{abstract}
Interactive intelligent systems, i.e., interactive systems that employ AI technologies, are currently present in many parts of our social, public and political life. An issue reoccurring often in the development of these systems is the question regarding the level of appropriate human and computer contributions. Engineers and designers lack a way of systematically defining and delimiting possible options for designing such systems in terms of levels of automation. In this paper, we propose, apply and reflect on a method for human---computer configuration design. It supports the systematic investigation of the design space for developing an interactive intelligent system. We illustrate our method with a use case in the context of collaborative ideation. Here, we developed a tool for information extraction from idea content. A challenge was to find the right level of algorithmic support, whereby the quality of the information extraction should be as high as possible, but, at the same time, the human effort should be low. Such contradicting goals are often an issue in system development; thus, our method proposed helped us to conceptualize and explore the design space. Based on a critical reflection on our method application, we want to offer a complementary perspective to the value-centered design of interactive intelligent systems. Our overarching goal is to contribute to the design of so-called hybrid systems where humans and computers are partners.
\end{abstract}

%
%

\ccsdesc[300]{Human-centered computing~HCI design and evaluation methods}
\ccsdesc[300]{Human-centered computing~Interactive systems and tools}

\keywords{Human---Computer Collaboration; Semantic Annotation; Large Scale Ideation}

\maketitle


\newcommand{\revision}[1]{{\textcolor{black}{#1}}}

\section{Introduction}
\label{sec:introduction}
From the beginning, the fields of human---computer interaction and computer-supported cooperative work have embraced the idea that human and machines, i.e., a computer, can engage in a productive relationship. Licklider's vision of a `man-machine symbiosis'~\cite{Licklider:1960th} and Engelbart's `augmenting human intellect' proposal~\cite{Engelbart:1962wn} have influenced how scholars have conceptualized this partnership profoundly. These ideas have been re-evaluated more recently because of the technological advancements. Regarding the concept of `human-computer integration'~\cite{Farooq:2016ey}, for example, the authors describe an interaction and integration continuum on which software can travel. Interactive intelligent systems are AI\footnote{The terms AI (artificial intelligence) or ML (machine learning) are subsequently deliberately omitted in this article, since the design approach proposed is not limited to a specific group of technologies.}-supported systems, with which people interact when selecting songs, reading news or searching for products. Such systems should be examined by considering both the human and the system. Jameson and Riedl call this a `binocular view' of interactive intelligent systems because the system's design includes algorithm design with interaction design, on the one hand, and a combined evaluation of a system's performance and human behavior, on the other hand~\cite{Jameson:2011fy}. However, existing research provides little guidance on how we should design interactive intelligent systems. Should a task be carried out by a human or a computer? What is an appropriate level of interaction vs. integration? Is human labor more preferable than automated action? How can we make an informed decision about allocating the task to either one or the other? How can we evaluate our decision? A proposal of a method for elaborating this spectrum is still missing.

The authors have experienced this issue in the context of information extraction from ideas in the research area of collaborative ideation. In addition to complete manual approaches for making sense of an idea's content~\cite{IBM2008}, algorithmic approaches were proposed which describe the content of the ideas statistically (e.g.,~\cite{Chan17}). However, both perspectives (manual vs. automatic) have their limitations; thus, the question is whether a `sweet spot' that emphasizes the advantages of both perspectives exists and how such a `sweet spot' can be identified. 
A `sweet spot' defines a compromise between the often contradictory evaluative criteria of an interactive vs. an intelligent system. \revision{It is a carefully balanced configuration of both parts, the intelligent and the interactive, by considering the objectives of the system's stakeholders and the context of use. The design of an interactive intelligent system might focus only on the learning part of the system, for example in the area of reinforcement learning~\cite{amershi2014power}. However, without considering the human perspective, that design might diminish the human's trust in that system and, therefore, the applicability of the system. In the context of our use case of collaborative ideation, the `sweet spot' defines the compromise between minimizing human effort in the annotation task (human perspective) and maximizing the necessary quality of the idea annotation (system perspective). A `sweet spot' represents one configuration of the interactive intelligent system, i.e. a specific task allocation between both partners.} Based on research from the area of human factors, we propose a method that allows for the configuration of an interactive intelligent system by defining and evaluating different levels of automation (LoA). Each configuration represents a certain LoA which involves a carefully designed technical realization (in terms of involved computation) and a specific amount of human involvement necessary. 

Our article makes the following contributions: (1) We propose a method for defining and evaluating different configurations of an interactive intelligent system; (2) we apply this method in the context of information extraction to the area of collaborative ideation; (3) we evaluate different configurations of information extraction empirically; (4) we provide a tool for information extraction that can be configured according to the specific goals of the use case; and (5) we offer a complementary perspective to the value-centered design of interactive intelligent systems.

We have organized our article into six parts. After this introduction, we introduce the related work which defines the context of our research. Building on that, we propose our model for the configuration of automation levels in interactive intelligent systems in the third part. We then introduce our use case on information extraction in the area of collaborative ideation and define suitable configurations. In the fifth part of the paper, we evaluate the different configurations empirically, examine the system and human perspective more closely, and discuss the lessons learned. In the last part of the paper, we reflect critically on our proposed method and provide some path for future research.


\begin{figure}[tb]
	\begin{center}
		\includegraphics[width=\columnwidth]{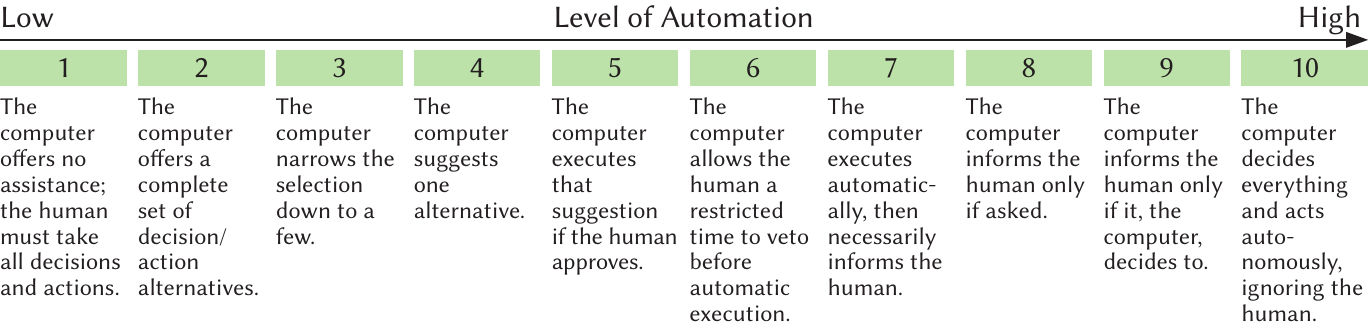}
		\caption{The ten Levels of Automation (LoA) showing possible interactions between human and computer organized by the amount of automation (adapted from Parasuraman and colleagues~\cite{Parasuraman:2000kv}).}
		\label{fig:loa_parasuraman}
	\end{center}
\end{figure}

\section{Related Work}
\label{sec:loa}

From the beginning, the development of software has been driven by the question: `What \textit{can} be automated?'~\cite{ arden1983can}. The goal was often to automate as many functions as technically possible~\cite{Dekker:2002fo}. However, instead of looking at these systems from a  machine-centered perspective, we should consider a human-centered perspective that changes the underlying question from `What can be automated?' to `What \textit{should} be automated?'~\cite{Tedre:2008ke}. We do not want to argue the ethical or philosophical part of this question, but rather ask for a methodical perspective on designing an interactive intelligent system that helps to answer this question explicitly. Even though 
Jameson and Riedl propose a `binocular view' of interactive intelligent systems that integrates algorithm design with interaction design and evaluates the system and the human integratively~\cite{Jameson:2011fy}, existing research provides little guidance on how such design and evaluation can be carried out in this context. 
We found this reoccurring question of automation, i.e., allocating tasks either to humans or to machines, in the area of robotics, ergonomics and human factors in our literature review. Following Parasuraman et al.~\cite{Parasuraman:2008iv}, `automation refers to the full or partial replacement of a function previously carried out by the human operator'. 

One of the first attempts to describe task allocation originates in Fitts's MABA-MABA (`Men are better at; Machines are better at') lists~\cite{Fitts:1951vb}. With the latter, Fitt aimed to support design decisions regarding whether a system's function should be carried out by either a machine or a human. The lists define a fixed set of skills and abilities that can be clearly attributed to either a human or a machine. However, these lists motivated one to compare human and machine capabilities instead of creating hybrid task allocations where human and machine activities intertwine; system engineers were thinking of replacing one with the other (mostly humans by computers). These led to a number of taxonomies --- so-called LoA --- with the first proposed by Sheridan and Verplank~\cite{Sheridan:1978ty}. They differentiate ten possible LoA in human---computer decision-making processes. The lowest levels describe a human who carries out the whole process up to the point when the computer takes over for realizing the decision. On the highest level, a computer carries out the task completely autonomously. In between, they define a continuum which details the different options of possible task allocation precisely. The research community on automation has reflected on this seminal work intensively (e.g.,~\cite{Endsley:1995bt},~\cite{Kaber:1997bz}) and this finally led to a taxonomy proposed by Parasuraman and colleagues~\cite{Parasuraman:2000kv}. This taxonomy (cf. Figure~\ref{fig:loa_parasuraman}) was accompanied by a framework for examining automation design issues for specific systems. It provides principled guidelines for automation design by defining a series of steps that are performed iteratively. The framework suggests four classes of functions: Information acquisition, information analysis, decision and action selection, and action implementation. Designers should evaluate possible LoA within each of these functional domains by studying the consequences of its associated human performance. Parasuraman and colleagues suggest a non-exhaustive set of evaluative criteria, such as reliability, operating cost and safety. Based on the evaluation, a system designer can recommend an appropriate upper and lower boundary on the level of automation that defines the maximal and minimal level required.

Dekker and Woods~\cite{Dekker:2002fo}, however, argue that these lists of automation levels convey the impression that such systems can be designed just `abracadabra.' A major issue is the underlying assumption that technology can replace humans. The goal of automation is conceptualized as a `substitution problem': A fixed workflow in which selected tasks are replaced (by automation), which leads to, amongst others, less labor, fewer errors and higher accuracy. Johnson and colleagues~\cite{Johnson:2017ep} suggest that the design of the `automated algorithm' process should be different for all joint human---computer activities which are not clearly separable functions. Thus, designers should support the interdependence of joint human---computer activities.


In addition to the traditional LoA approach, a number of similar approaches have been suggested to design human---machine task allocation. One alternative discussed widely is the \textit{mixed-initiative interaction} approach~\cite{Horvitz:1999mi}, in which a human---machine activity is a joint activity where human and machine interact and negotiate their actions. Fong proposes a similar approach~\cite{Fong:2001vg} and defines the concept of \textit{collaborative control}, describing the simultaneous involvement of human and machines in the same activity. The work on \textit{coactive design} proposed by Johnson and colleagues~\cite{Johnson:2011jl} goes in a similar direction. They argue that instead of allocating functions, the focus should be on supporting interdependency. All these approaches have in common that humans and machines complement rather than replace each other~\cite{Jordan:1963va}. Another line of research focuses on the question how human computation can be flexibly combined with algorithms to create new hybrid human---machine-based algorithms~\cite{Demartini:2017ey, Kamar:2012vb}.

Isbell and Pierce~\cite{Isbell:2005vh} translated the LoA into the context of human---computer interaction. 
If the computer offers no assistance, researchers are concerned mainly with questions in the area of direct manipulation. On the other end of the spectrum, research originates that focuses on intelligent user interfaces.
Both poles represent the seminal discussion between Ben Shneiderman and Patti Maes~\cite{Shneiderman:1997rw}, which conveyed this idea that software agents\footnote{One might think of `Clippy' when reading the word `software agent,' but we relate more to adaptive functions.} should not be seen as an alternative to direct manipulation, instead, they are complementary metaphors.
In all cases, a well-designed user interface is necessary for a software agent to interact with a human. 
This complementary perspective is reflected by the proposal of a concept by Farooq and Grudin~\cite{Farooq:2016ey}, who define their continuum from interaction to integration and call for more integrative design and evaluation approaches.
Jameson and Riedl propose a `binocular view' of interactive intelligent systems that integrates the algorithm design with the interaction design in terms of building and evaluating the system~\cite{Jameson:2011fy}. 

However, except for the framework of Parasuraman and colleagues~\cite{Parasuraman:2000kv}, all approaches mentioned provide no clear guidance for designing interactive intelligent systems. Thus, the LoA framework finally informed our proposal of a method for designing and evaluating the human---machine configurations which we introduce in the next section.

\begin{figure}[bt]
	\begin{center}
		\includegraphics{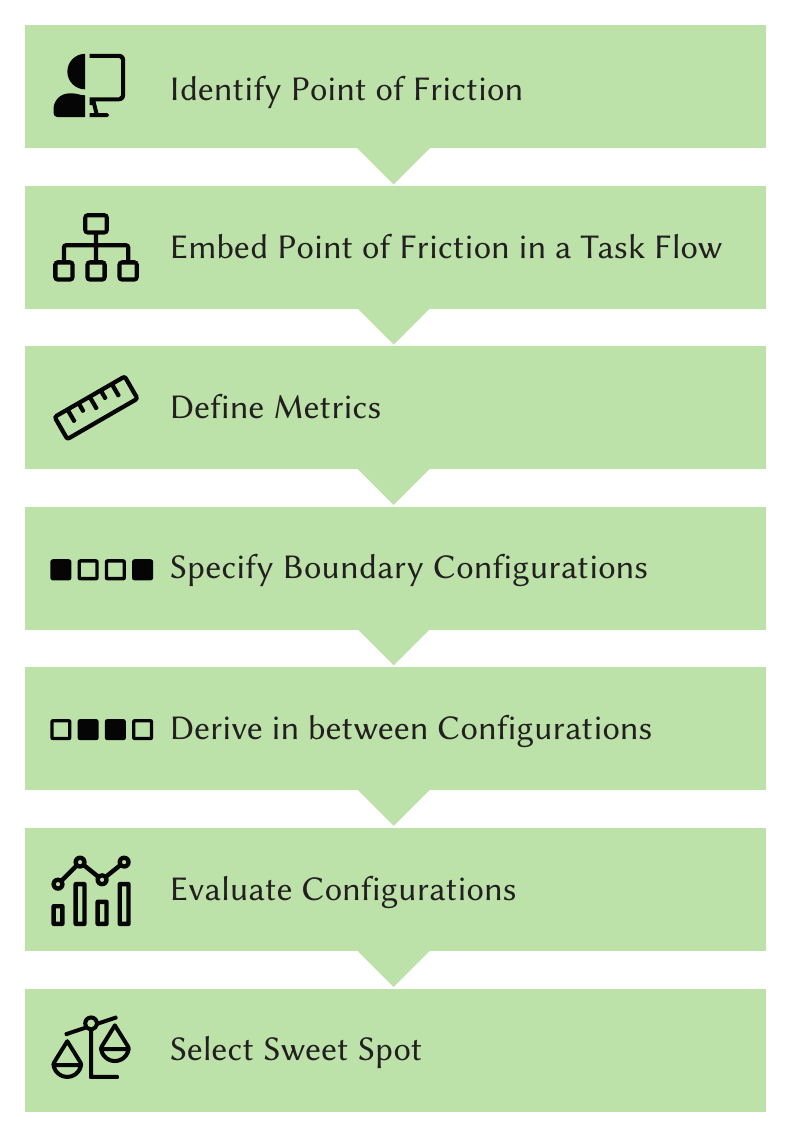}
		\caption{Stages of the proposed method for designing configuration levels of interactive intelligent systems. The different stages are carried out iteratively. During each iteration, a new increment, i.e., configuration, is defined, adapted or evaluated. }
		\label{fig:method}
	\end{center}
\end{figure}

\section{Method for Human---Computer Configuration Design}
\label{sec:method}

We transferred the framework proposed by Parasuraman and colleagues~\cite{Parasuraman:2000kv} from the design of an automated system with a fixed level of automation to the design of interactive intelligent systems with different configurations. 
Our perspective differs from the work discussed insofar as we want to allow for the configuration of an interactive intelligent system according to the specific requirements of an application domain and the system's stakeholders. \revision{What do we mean by configuration? 
	Instead of implementing a system with a fixed set of functions, i.e., a fixed level of automation, the design and the deployment of an interactive intelligent system should follow a more flexible approach. Each application domain requires a calibration of the system, since the context and, therefore, the requirements change. Such calibration is carried out by specifying or selecting possible configurations which are derived from the LoA. All configurations represent the design space of possible task allocations in interactive intelligent systems which need to be evaluated for each context of use.}

\revision{We defined the proposed method in an incremental and iterative manner. We started with a draft description that followed existing work quite closely~\cite{Parasuraman:2000kv}. Based on our experiences during the application of the method in our use case, we refined each stage more precisely if necessary. The original model is based on the area of automation; thus, the scope is much broader. We focus on the usage of intelligent systems, which can be defined as systems that employ AI technologies, for example, in recommender systems, expert systems, knowledge-based systems or as intelligent agents. We narrowed the focus of the proposed method, based primarily on our own experiences, down to decision-support systems. This led finally to a proposal of a configuration design method that consists of seven stages (cf. Figure~\ref{fig:method}). Even if we present these individual stages in a sequence, stages 2 to 5 especially should be carried out iteratively instead. We briefly describe each stage in the next sections. We then apply the proposed method on the use case from the area of collaborative ideation.}

\subsection{Identify the Point of Friction}
\label{sub:identify-point-of-friction}
\revision{In the first stage, we identify the part of the interactive intelligent system in which the system's objectives collide with the human's objectives --- the point of friction, i.e., where the specific goals of the human and the computer diverge.  
	Parasuraman and colleagues suggested four components for separating more complex human---computer workflows into main processes:
	Information acquisition, information analysis, decision and action selection, and action implementation~\cite{Parasuraman:2000kv}.
	This distinction can provide an initial indication for identifying such points of friction. However, a specific concern often exists implicitly when designing an interactive intelligent system, since it might not be apparent how much and which form of automation might be required. Possible contradicting objectives can be disclosed by identifying the different objectives of the stakeholders of a system. This stage, as we discuss later in this article, makes this research particularly compatible to the value-sensitive design approach~\cite{Friedman:1996gc, Friedman:2017eu}, in which the stakeholder analysis and the identification of possible tensions between existing values are firmly anchored in the methodology.}

\subsection{Delimit the Point of Friction in a Task Flow}
\revision{In the second stage, we translate the point of friction into a model (e.g., task flow) which describes the sub-tasks within the process of human---computer integration. During that stage, a neutral perspective in terms of the allocation of the sub-task to either a human or a computer is required; therefore, we recommend defining a task flow either consistently from one perspective (human or computer) only or without incorporating the execution. A challenge in this task is the `right' level of modeling the process. A task flow model which is too abstract might hide possible task allocations (too simple design space), whereas a model which is too detailed might reveal too many possible task allocations which could lead to too many possible configurations (too complex design space). Thus, modeling the task flow is closely linked to the step of deriving the in-between configurations.}

\subsection{Define Metrics for Evaluating Configurations}
\label{sub:method_metrics}
\revision{After modeling the task flow, we need to elaborate on the point of friction between the interactive and the intelligent system. The underlying question is: How can we evaluate the effectiveness of the different configurations? By looking at existing research, we can find a variety of measures that allow assessing the effectiveness of an interactive (e.g.,~\cite{olson2014ways}) or intelligent system (e.g.,~\cite{precrec79}).  
	On the one hand an interactive system is primarily evaluated with a human in mind with a focus on interaction. On the other hand, the intelligent system is, instead, evaluated with a focus on the technical realization, for example, looking at the efficiency and correctness of the algorithm implemented. Thus, we propose to define metrics that consider either only the human or the algorithmic perceptive. This understanding allows us to conceptualize the point of friction as a trade-off between the social and the technical side. \\
	However, by looking at interactive intelligent systems, we realize that such separation simplifies the whole system. One might ask whether we can capture all characteristics of a configuration by this conceptual separation. Various references in the literature suggest the opposite. Winner describes, based on diverse examples, the influence of technology on specific societal groups and highlights the political character of technical decisions~\cite{Winner:1980vp}. By incorporating this argument, Latour argues that the delegation of tasks previously carried out by people to technologies (or vice versa) can lead to significant shifts in social practices and responsibilities~\cite{Latour:1988co}. Not only sociologists but also computer scientists advocate similar viewpoints. Dekker and Woods state that replacing or extending a human responsibility by an algorithmic component cannot be captured by only one or the other perspective since such hybrid workflows cause a qualitative change regarding how people perceive a technical system~\cite{Dekker:2002fo}. Jameson and Riedl define this as a `binocular view' of interactive intelligent systems which call for studies that investigate the performance of algorithms and human behaviors side by side~\cite{Jameson:2011fy}. Previous work seems to emphasize the need for ethnographic studies which allow capturing  all the little nuances of technology usage --- the qualitative shifts. However, such ethnographic research is rather appropriate in the deployment phase of a system, while we focus on the design phase. We suggest, therefore, identifying a heuristic that allows the capture of the attitude of humans toward the intelligent system. In summary, we propose metrics that will enable the evaluation of the different configurations from a human (interactive) and a computer (algorithmic) perspective but also both an integrative human---algorithm view.} 

\subsection{Specify Boundary Configurations}
The next two stages deal with the actual implementation of the configurations. At first, we define the two `extreme configurations' which are the norm in other settings (interactive/intelligent system design). These so-called boundary configurations result in the outer edges of the instantiated LoA spectrum. One side describes a completely manual realization of the task flow (\textit{all-human}), whereas the opposite side exhibits the fully automatic configuration (\textit{all-computer}). We assume that a point of friction exists in our context that hinders the realization of one or the other configuration alone. Thus, we consider the `sweet spot' always originates in between these two edges. However, there might be use cases in which the previous elaborations have already revealed that one or the other boundary configuration might not be desirable. In such cases, a lower or upper level of the LoA depicts the edges of the spectrum. We next explain how we identify the different configurations in this spectrum.

\subsection{Derive In-between Configurations}
\revision{
	The edges of the spectrum defined by the LoA are the starting point for adapting the task flow in terms of different configurations.
	We recommend analyzing each sub-task from both a human and computer perspective. The remaining eight LoA (cf. Figure \ref{fig:loa_parasuraman}, ~\cite{Parasuraman:2000kv}) guide the identification of possible interactions between human and computer.
	The fourth level, for example, states: 
	\textit{'The computer suggests one alternative.'} which can be interpreted as an explicit instruction for the technical realization of the configuration.
	Within a recommender system, for example, the system provides one recommendation only, which can either be accepted by a person or not.
	To specify a configuration, we suggest that each level is examined carefully and then this level is superimposed with the sub-task, whether a human, a computer, or both can fulfill it.
	For those sub-tasks where both --- a human and a computer --- might be involved, we suggest splitting up the task even further until a precise distinction is possible.
	Finally, each of the configurations defined is implemented prototypically.
}

\subsection{Evaluate the Configurations Based on Metrics}
\revision{In the second to last stage, the various configurations are assessed based on the metrics defined. These metrics already inform the study design, since the complete range of methods is theoretically possible. In one use case, for example, a decision-support system in an employment office needs to be realized. The system's designers have access to various stakeholders (e.g., employees, manager, potential applications, workers union); thus,  these stakeholder groups can evaluate the different configurations in close feedback circles. In other cases, online experiments in the form of crowd-sourcing studies might be more applicable. A purely technical evaluation of the interactive intelligent system is less appropriate, since it undermines the socio-technical perspective of this work.}

\subsection{Decide on the Sweet Spot Configuration}
\revision{Finally, based on the evaluation results, the actual exploration of possible `sweet spots' takes place, and an informed decision is possible for choosing the most suitable configuration for the interactive intelligent system. The `sweet spot' defines the configuration where the point of friction between the human and the computer is minimal. In the end, the designer and the different stakeholder groups must decide on which configuration is best-suited; thus, this decision is highly dependent on both the context and the specified metrics. 
	The values obtained for the metrics in each configuration provide an explicit operationalization of the trade-offs defined in the point of friction. Thus, they allow for the comparison of the  benefits and drawbacks in each configuration and a subsequent informed decision on a `sweet spot' configuration. 
}
In the following, we apply this method to a use case from the research area of collaborative ideation. After introducing the context of this project and our motivation for designing an interactive intelligent system, we apply each of the stages described to the project.

\section{Use Case: Collaborative Ideation}
\label{sec:icv}
Collaborative ideation platforms, which rely on large numbers of ideators who generate many ideas in a distributed setting, have emerged as a promising solution to obtain different ideas from the crowd.
However, the practice of such crowd-based ideation has revealed a significant limitation: Many ideas have a mundane, rudimentary and repetitive character~\cite{siangliulue2015toward,siangliulue2016ideahound}.
Therefore, creativity-enhancing interventions were proposed as approaches to improve the quality of the ideas provided~\cite{siangliulue2015providing,siangliulue2015toward,siangliulue2016ideahound,girotto2017effect,Chan17}.
A prerequisite for providing and evaluating creativity-enhancing interventions is an understanding of the meaning of ideas created by people.
However, a manual evaluation of ideas does not scale well.
The assessment of 46,000 ideas submitted to the IBM 2006 innovation jam, for example, took reviewers six months~\cite{IBM2008}.
Thus, scholars use techniques from AI to reveal the meaning of ideas automatically.
Some work employs statistical models that define the similarity between words in ideas by determining the closeness of these words in a multi-dimensional space. The closer the words, the more similar they are~\cite{Chan17}.

Other work uses idea annotation as an approach to extract information about ideas.
This means that idea texts are enhanced by annotations that link terms used in an idea (a term can be a single word, e.g., \textit{door}, or a compound word, e.g., \textit{pet food}) with concepts from an external data source.
Popular approaches utilize the lexical database WordNet~\cite{martinez2018information}.
In the latter, lexical relationships between concepts are modelled, for instance, `flower' is a `plant' and a `plant' is an `organism.'
Other research uses knowledge graphs for enriching ideas semantically~\cite{gilon2018analogy}.
A knowledge graph organizes various topically different real-world entities, called concepts, with their relationships in a graph.
It also provides a schema that aggregates these concepts into classes (abstract concepts) which also have relationships with each other~\cite{paulheim2017}.
Knowledge graphs describe the similarity between concepts by a richer set of relations (e.g., `is-a-kind-of,' `is-a-specific-example-of,' `is-a-part-of,' `is-the-opposite-of').
An example for a knowledge graph is DBpedia\footnote{Further information can be obtained via \url{https://wiki.dbpedia.org/}.}, which is built by extracting structured data from Wikipedia.

The existing relationships between concepts in a knowledge graph allow us to identify similar ideas, even though they have no terms in common, by revealing a relationship on a more abstract level, for example, between abstract concepts.
An idea describing modification of a door and another describing wall painting, for example, could be connected by the concept of \textit{architecture}. Thus, as opposed to statistical methods that are based on explicit relationships, knowledge graphs allow for identifying implicit, more subtle relationships.

Enabling the annotation of concepts in idea texts would have multiple advantages.
The concepts could be used, for example, as an input for a faceted search tool (e.g., finding all ideas that talk about kinds of plants), they enable aggregation and cluster visualization of the solution space~\cite{yoo2015ontology}, and they could potentially enable effective inspiration mechanisms such as \textit{analogical transfer}~\cite{gilon2018analogy} (via a `has-usage' relationship between concepts).

This potential of annotating ideas motivated the development of our `Interactive Concept Validation' (ICV) technique, which can be integrated into traditional ideation processes \cite{mackeprang2018concept}. After submitting an idea, a person is asked to annotate the idea manually based on concepts obtained from a knowledge graph. This tool consists of an interactive component, since we ask a person for its annotations and an intelligent component, because the provision of the annotations uses AI technologies.

\begin{figure}[t]
	\begin{center}
		\includegraphics[width=\columnwidth]{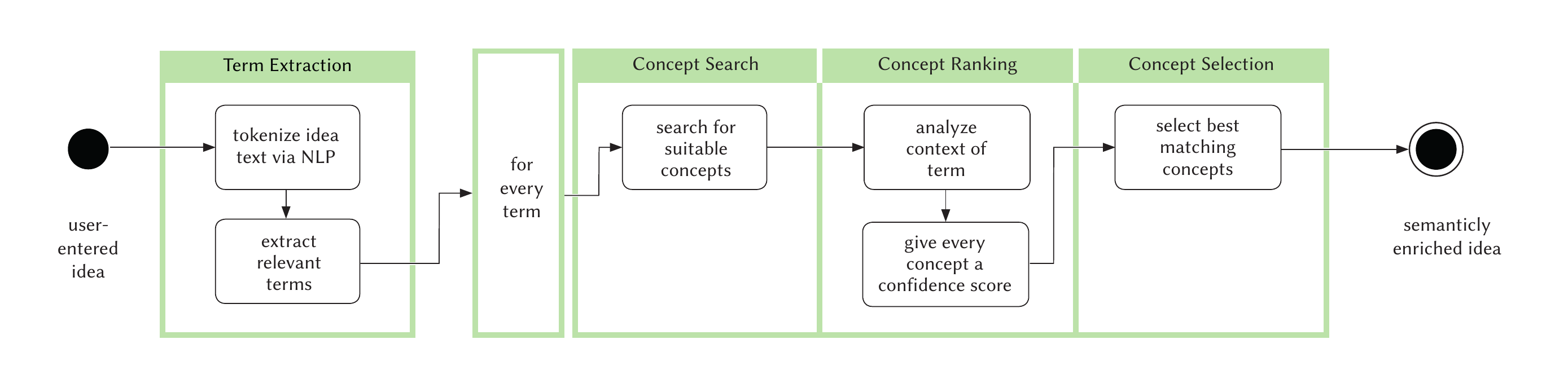}
		\caption{Task Flow Diagram of sub-tasks identified in the interactive concept validation (ICV) process.}
		\label{fig:processOverview}
	\end{center}
\end{figure}

\subsection{Identify Point of Friction}
\label{sec:icv_pointsoffriction}

We identified two opposite approaches while designing the software that enriches ideas semantically.
On the one hand, manual approaches build tools for domain experts who can select text manually and connect it to a knowledge graph~\cite{breitenfeld2018enabling,grassi2013pundit,widlocher2012glozz}.
On the other hand, \textit{automatic concept extraction and linking} uses algorithmic approaches to label text automatically~\cite{isem2013daiber}.
Manual approaches do not scale well, whereas automatic approaches have to deal with the issue of \textit{word-sense disambiguation}.
The latter means that the term `keyboard,' for example, could refer to a typewriter keyboard (computer technology) or the musical instrument.
For humans, the meaning can be easily derived from the context, for algorithms, this is still challenging if not impossible. 
Therefore, automatic semantic enrichment of ideas should benefit greatly from human contribution.
However, ideation is a task that should only be interrupted as little as possible, as we assume that distraction could impede creativity. 
This point of friction emerged during software development. It defines a trade-off between a human perspective that requires maximal automation of the idea annotation to distract the ideation process as little as possible, and a data perspective that requires a high quality of annotated ideas. 

We decided on the level of automation without reflecting on this point of friction in an initial implementation of the software. However, when discussing the results of this initial implementation, we realized the random character of our design decisions. Thus, the concept extraction from ideas is an ideal candidate for applying our method of discovering the `sweet spot' between human and algorithmic contributions. Our goal is to make a more informed decision regarding the question of the automation needed and required in this context. Starting from these considerations, we enter the second stage of our method by looking more closely into the conceptual realization of this information extraction task. 

\subsection{Delimit the Point of Friction in a Task Flow}
\label{sec:task-modeling}
The ICV technique consists of one primary task which represents the existing point of friction: We ask a person to disambiguate all terms in their generated ideas by providing possible meanings for each term (if there are more than one).
A person can select a concept that represents the meaning of the term best. We use concepts from knowledge graphs for providing the possible meaning of terms. We detail this primary task more precisely in the following. 

We visualized the task in a task flow diagram in Figure~\ref{fig:processOverview}. 
The ICV process starts after a person creates an idea.
Each idea consists of a short text in which all terms (words and compound words) need to be identified. This step is called \textit{Term Extraction}.
In a next step, all extracted terms in an idea are searched for suitable concepts within a knowledge graph. We use the knowledge graph DBpedia.
All possible concepts then have to be ordered by their suitability to describe the term under focus. This step is called \textit{Concept Ranking}. 
These ranked concepts need to be reviewed to select the most suitable concept in the \textit{Concept Selection} step.
The result of the ICV is the semantically enriched idea, i.e., the idea text with a set of annotated terms. Each annotated term links to a concept in the knowledge graph, which allows the meaning of ideas to be captured in more detail. 

\subsection{Define Metrics for Evaluating Configurations}
\label{sec:metrics}

After modelling the task flow, we conceptualize the trade-off described by defining metrics that take the point of view of either the human (by favoring an interactive software) or the algorithm (by focusing on automating the task as much as possible).

The overarching goal of the ICV is to provide precisely annotated ideas that can be used to support the collaborative ideation process in later stages, for example, by providing suitable creativity-enhancing interventions. Based on the discussion in Section~\ref{sub:method_metrics}, we discuss possible metrics from three perspectives: The human, the algorithmic and both. 

\subsubsection{Human Perspective}
\label{sec:metricsHumanPerspective}
From a human perspective, our goal is to minimize the effort of the person involved.
By doing so, we want to ensure that the ICV is as noninvasive as possible regarding the ideation process.
We can capture the effort by collecting data on human activities or asking a person to rate their perceived effort. 
Regarding data collection, we can measure the \textit{interaction effort} of a person by instrumenting the software with a tracking system which records the number of \textit{clicks} and the \textit{time} needed to perform the ICV.
A widely used approach to capture mental effort is the \textit{Nasa Task Load Index (TLX)} questionnaire~\cite{hart1988development}.
We employed the `Raw TLX' by asking an ideator for their perceived mental and temporal demand, combined with their self-assessment of performance, effort and frustration during the ICV task\footnote{As the TLX physical demand scale refers to physical activities (e.g., pushing, pulling), we excluded it from our evaluation.}.

\revision{Furthermore, the influence of the ICV on the human ideation processes (their creativity) also needs to be captured. 
	The operationalization of creativity in collaborative ideation is most often based on two metrics: Fluency~\cite{siangliulue2015providing} and dissimilarity of ideas~\cite{girotto2017effect}. Fluency describes the number of ideas produced by a participant. The dissimilarity of ideas can be computationally evaluated by the deployment of a text-similarity algorithm (e.g., Latent Semantic Analysis) and subsequent determination of the two most distant ideas for a specific user.}


\subsubsection{Algorithmic Perspective}
\label{sec:metricsAlgorithmicPerspective}

From an algorithmic perspective, we need to measure the quality of the information extraction task, i.e., the suitability of the concepts annotated in describing the terms.
An established way of measuring the quality in the area of information extraction is to measure \textit{precision}, \textit{recall} and the \textit{F-measure}.
Whereas \textit{precision} defines the number of concepts annotated correctly, \textit{recall} defines the number of concepts correctly disambiguated relative to the number of all concepts found.
The \textit{F-measure} is the harmonic mean of precision and recall.
A prerequisite for using these measures is a so-called gold standard; in our context, it means that we need to create a manually annotated idea corpus.
We explain the creation of such corpus in more detail in Section~\ref{sec:boundaries-specification} and its usage in Section~\ref{sec:boundaries-evaluation}.

\subsubsection{Human---Algorithmic Perspective}
\label{sec:metricsHuman+AlgPerspective}

As stated in Section~\ref{sub:method_metrics}, we assume that an interactive intelligent system, such as our ICV software, might influence how people experience the ideation process itself. Thus, we need an approach that helps us to understand how humans experience the software. Such a gain in understanding can be reached by asking people how they perceive the different configurations from various angles. 

We decided to use the recommender systems' quality of user experience (ResQue) questionnaire, which captures acceptance, usability and satisfaction based on the user's perception and not the quality of the algorithm~\cite{pu2011user}. We found this questionnaire especially suitable for three reasons.

Firstly, the ICV can be defined as a typical recommendation process. A person gets recommendations, i.e., concepts, and selects the best fitting concept from the list of recommendations. 
Secondly, in contrast to questionnaires such as TAM\footnote{TAM stands for \textit{Technology Acceptance Model}. For further information we refer to Venkatesh \& Bala~\cite{venkatesh2008tam}.}, the ResQue questionnaire offers validated questions for our context of use --- recommender systems. 
Finally, as opposed to SUMI\footnote{SUMI stands for \textit{Software Usability Measurement Inventory}. Further information is available at~\url{http://sumi.uxp.ie}.}, the ResQue questions are designed to be adapted for the purpose at hand and offer insights with fewer items (11 vs. 50).
The ResQue questionnaire is based on four dimensions: \textit{Perceived system qualities}, \textit{user beliefs}, \textit{user attitudes} and \textit{behavioral intentions}. As behavioral intention captures attributes of recommender systems geared towards product recommendations, we excluded this dimension from our metric set. This left us with the 11 questions used in the study (cf. Appendix~\ref{app:resque} for the dimensions and questions).



\subsection{Specify Boundary Configurations}
\label{sec:boundaries-specification}

In this stage, we define the boundary configurations that depict the human vs. algorithmic realization of the annotation task. On an abstract level, the boundary configurations of our use case represent the `extreme configurations' for an information extraction task as described in Section~\ref{sec:icv_pointsoffriction}. The manual configuration represents the complete manual annotation of ideas based on human labor. We have already argued that such an approach is not scalable; however, we assume that manual annotation leads to the highest data quality because of the challenge of word-sense disambiguation. We use the results of the completely manual annotation of the ideas as a gold standard for evaluating the quality of the annotation in the other configurations.
Algorithmic approaches of text enrichment are situated at the other end of the spectrum that allow one to carry out all the sub-tasks defined (cf. Figure~\ref{fig:processOverview}) automatically. As \textit{automatic concept extraction and linking} is a flourishing field of research, several systems are suitable for such automation. The defined boundary configuration serves as the outer limit of our design space.

\begin{figure}
	\centering
	\includegraphics{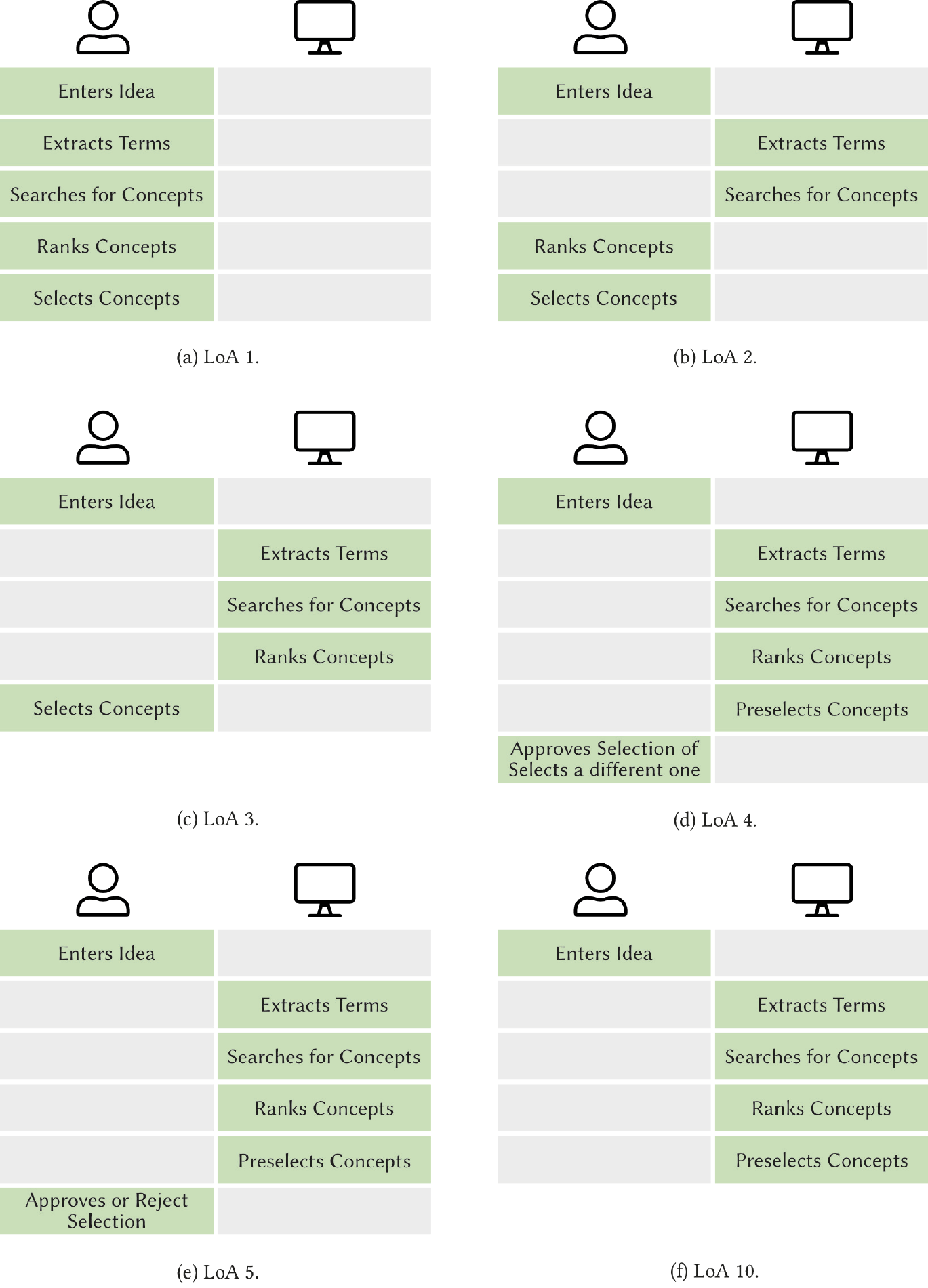}
	\caption{Different Levels of Automation in the ICV Process.}
	\label{fig:loa}
\end{figure}

\subsection{Derive In-Between Configurations}
\label{sec:applied-loa}

\revision{In the next step, we used the definitions of the remaining LoA to specify possible system configurations for our task flow.
	As the ICV process starts after a person has created an idea, we employed the `human enters idea' sub-task as the starting point for all task configurations. Based on the detailed description of the LoA proposed by Sheridan and Verplank~\cite{Sheridan:1978ty} and the taxonomy proposed by Parasuraman and colleagues~\cite{Parasuraman:2000kv}, we detailed each of the configurations by allocating the task to one or the other partner.} 

\revision{The boundary configurations (as shown in Figure~\ref{fig:loa}a and ~\ref{fig:loa}f) define the ends of the spectrum by an allocation of all sub-tasks to either a human or a computer. Based on that, we systematically went through the range from the lowest non-boundary level (LoA 2) to the highest (LoA 9).}

We allocated the sub-tasks `extract terms' and `search for concepts' to the computer for LoA 2 (cf. Figure~\ref{fig:loa}b). These sub-tasks are linked: We need to `extract the terms' from the ideas to `search for concepts' in the knowledge base. 
For LoA 3, we interpreted the \textit{narrowing down} of the selection as an automatic ranking of concepts done by the computer. When moving towards further automation, we defined a task allocation that enables the computer to either preselect concepts (LoA 4, cf. Figure~\ref{fig:loa}d) or select them automatically (LoA 5, cf. Figure~\ref{fig:loa}e).
When discussing the levels 6 to 9 of the model defined by Parasuraman, we detected a point of uncertainty for our use case. While the levels define the interaction between human and computer in terms of decision-making, there is no definition given about the possibility of changing the results afterward. After discussing possible configurations, we decided to exclude the levels 6 to 9 from further evaluation, as changing the results of the concepts provided did not seem suitable for our use case.

In summary, Table~\ref{tab:HM_ICV} describes the task allocation for human, computer or both for each configuration. We used each configuration to develop a software prototype (see Section~\ref{sec:EvalutingInBetweenConfigurations}) which encodes the configuration, and we evaluated that subsequently.

\begin{table}[t]
	\begin{tabularx}{\textwidth}{XC{0.5cm}C{2cm}C{2cm}C{2cm}C{2cm}}
		\toprule
		Configuration & LoA & Term Extraction & Concept Search & Concept Ranking & Concept Selection \\ \midrule
		All-Human & 1 & \faUserO & \faUserO & \faUserO & \faUserO \\
		Baseline & 2 & \faDesktopS & \faDesktopS & \faUserO & \faUserO \\
		Ranking & 3 & \faDesktopS & \faDesktopS & \faUserO~\faDesktopS & \faUserO \\
		Validated Threshold & 4 & \faDesktopS & \faDesktopS & \faDesktopS & \faUserO~\faDesktopS \\
		Automated Threshold & 5 & \faDesktopS & \faDesktopS & \faDesktopS & \faUserO~\faDesktopS \\
		Ranking/A. Threshold & 5 & \faDesktopS & \faDesktopS & \faDesktopS & \faUserO~\faDesktopS \\
		All-Computer & 10 & \faDesktopS & \faDesktopS & \faDesktopS & \faDesktopS \\ \bottomrule
	\end{tabularx}
	\caption{Task allocation table for different LoA: Each sub-task for each configuration is allocated to either the human (\protect\faUserOSmall), the computer (\protect\faDesktopSmall) or both (\protect\faUserOSmall~\protect\faDesktopSmall).}
	\label{tab:HM_ICV}
	\label{results-FBD}
\end{table}


\section{Evaluate Configurations based on Metrics}
After modeling the task flow, specifying metrics and applying automation levels, the next step was to evaluate the different automation configurations in experimental settings.
In the following, we describe how we evaluated boundary configurations of our human---computer spectrum and then the configurations in between.

\subsection{Evaluating Boundary Configurations}
\label{sec:boundaries-evaluation}
After the specification of boundary configurations in Section \ref{sec:boundaries-specification}, we implemented both: We defined an annotation scheme to create a gold standard of annotations for the manual side; we employed a software for concept extraction and linking for the fully automated configuration.

\subsubsection{Realization of the Manual Boundary Configuration}
\label{sec:manualedge}
In our context, we defined the manually annotated gold standard by sampling 60 idea texts obtained in previous studies\footnote{For the published idea data sets, see \url{https://osf.io/k2ey7/}.}, and then annotated them using the method described subsequently.
We (two of the authors) annotated all ideas based on a three-round process, whereby we carried out each step in the task flow described in Figure~\ref{fig:processOverview} manually.
In the first round, we selected five ideas randomly and annotated them (from the task term extraction to concept selection) collaboratively.
At first, we read each idea and chose appropriate terms as concept candidates.
We then searched for potentially fitting concepts on the knowledge graph DBpedia by using its web user interface\footnote{The interface is available at \url{http://dbpedia.org/fct/}.}.
We finally selected the most appropriate concepts by discussing all the concepts obtained.
We proceeded with this process until we had annotated all possible terms in all five ideas.
The first round in the manual process allowed us to develop a shared understanding of the information extraction process.
We, therefore, carried out the second round (the remaining 55 ideas) independently and compared our results (Cohen's kappa $0.68~\%$ \cite{cohen1960coefficient}).
In the third round, we checked all concepts identified per idea and discussed all ambiguous cases.
If necessary, we adapted the concepts and cross-validated all concepts over the entire idea set.
The resulting data set contained 281 concepts for all 60 ideas.
We needed an average of about 20 minutes to annotate and discuss each idea, which resulted in about 20 hours of human labor for the task.

\begin{figure*}[tb]
	\centering
	\savebox{\imagebox}{\includegraphics[width=0.441\textwidth]{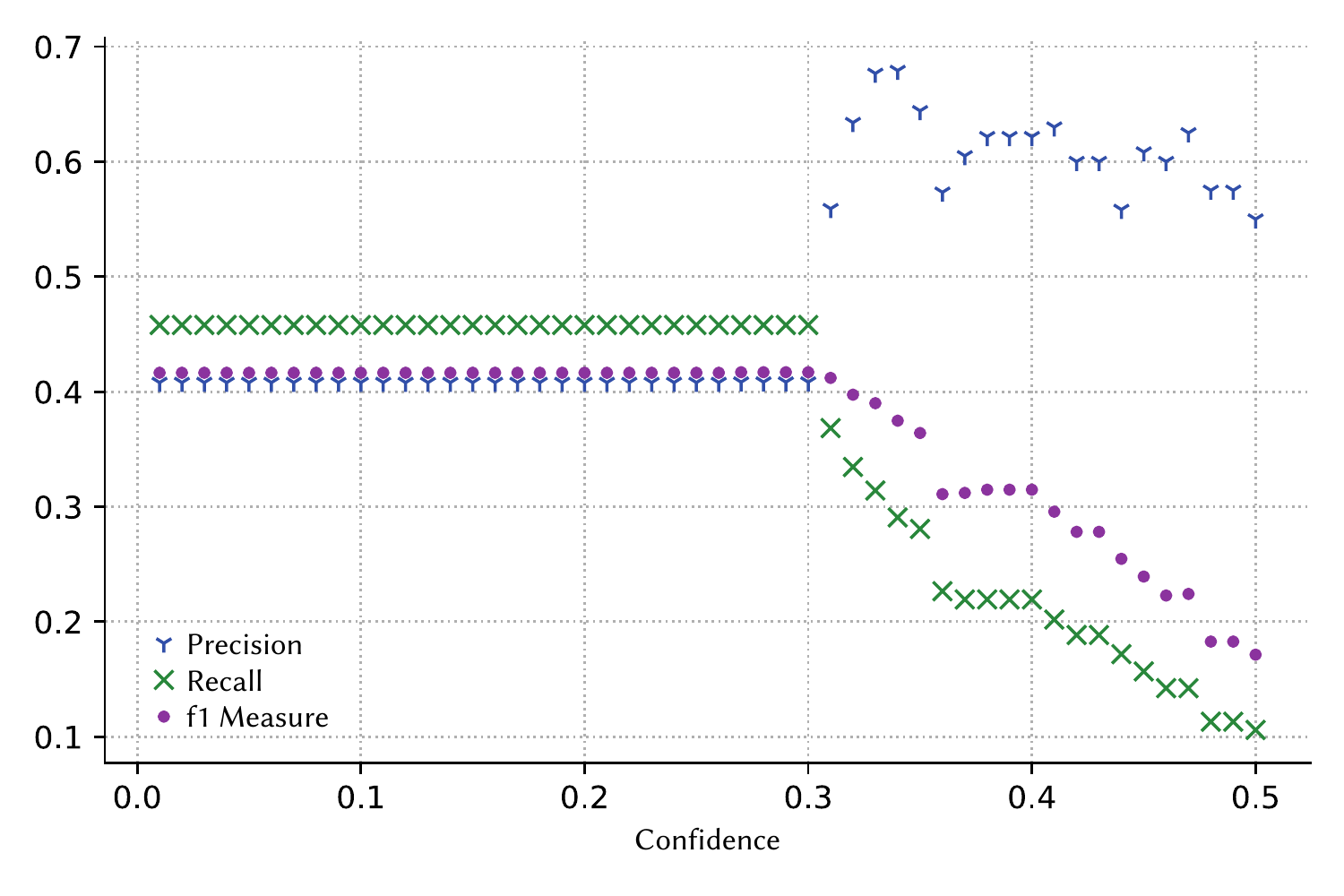}}%
	\begin{subfigure}[b]{0.49\textwidth}
		\centering\usebox{\imagebox}
		\caption{Results for precision, recall and F-measure for different spotlight threshold parameters based on the sample idea data set.}
		\label{fig:precRecCSCW19-1-sample}
	\end{subfigure} \hfill 
	\begin{subfigure}[b]{0.49\textwidth}
		\raisebox{\dimexpr.357\ht\imagebox-.0\height}{
			\resizebox{0.9\textwidth}{!}{%
				\begin{tabular}{lccc}
					\toprule
					Confidence & Precision & Recall & F-Measure \\ \midrule
					\textbf{$\gamma=0.1$} & 0.41 & \cellcolor{gray!25} 0.46 & \cellcolor{gray!25} 0.42 \\
					\textbf{$\gamma=0.3$} & 0.41 & \cellcolor{gray!25} 0.46 & \cellcolor{gray!25} 0.42 \\
					\textbf{$\gamma=0.4$} & \cellcolor{gray!25} 0.62 & 0.22 & 0.31 \\ \bottomrule
				\end{tabular}%
			}
		}
		\caption{Results showing the precision, recall and F-measure of the automatic concept extraction system for the data set with different confidence thresholds.}
		\label{tab:results-automated-annotation1}
	\end{subfigure}
	\caption{Data quality results for fully automatic information extraction by using DBpedia Spotlight on the 20 ideas.}
	\label{fig:precRecResults}
\end{figure*}

\subsubsection{Realization of the Automatic Boundary Configuration}
\label{sec:automaticedge}

We carried out a literature study and carefully read existing software comparisons in the field of algorithmic information extraction~\cite{martinez2018information}.
We decided to use the software DBpedia Spotlight as the fully automatic annotation system~\cite{Mendes:2011}.
This software allows for automating the whole process of annotation, as shown in Figure~\ref{fig:processOverview}.

Spotlight uses a three-step process of \textit{spotting}, \textit{candidate selection} and \textit{disambiguation}.
At first, Spotlight identifies all terms in an idea that possibly relate to existing concepts in the knowledge graph DBpedia.
It then selects \textit{concept candidates} that could describe that term by using the \textit{DBpedia Lexicalization dataset}\footnote{The DBpedia Lexicalizations Dataset stores the relationships between DBpedia resources and a set of terms that are potentially linked to those resources.}.
Spotlight then analyzes the context of the current term and selects (or \textit{disambiguates}) the best fitting concept.
A core component of the Spotlight software is the possibility of tuning the disambiguation step to improve the results for a specific use case.
The most promising tuning option for us was the \textit{disambiguation confidence}, for which a sample of 100,000 Wikipedia articles was used to determine it~\cite{Mendes:2011}. The threshold ranges from 0 to 1 and offers a percentage of certainty the spotlight algorithm has in disambiguating every given term. This uncertainty is defined as follows: ``a confidence [threshold] of 0.7 will eliminate 70~\% of incorrectly disambiguated test cases.''
We aimed at evaluating the result of the fully automatic information extraction task by the resulting quality of the annotations (cf. discussion in Section~\ref{sec:metricsAlgorithmicPerspective}).
We computed the precision, the recall and the F-measure against the manually annotated idea set by using different confidence thresholds\footnote{We used the Spotlight Web API available under~\url{http://api.dbpedia-spotlight.org/en/annotate} and confidence threshold steps of $0.01$.}.
For this, we used the manually annotated idea corpus (cf. Section~\ref{sec:manualedge}) as a gold standard. 

The results are visualized in Figure~\ref{fig:precRecCSCW19-1-sample}.
The values for all three metrics for our data set remained relatively stable for confidence thresholds under $0.3$.
When the confidence threshold was higher than $0.3$, we observed a decline of recall and F-measure. 
We decided to use three confidence thresholds, namely, $\gamma=0.1$, $0.3$ and $0.4$, for the fully automatic information extraction task to capture the different quality levels with the automated task realization. 
A request to the Spotlight API took us about one second, which resulted in a time effort of 60 seconds per confidence threshold.

\subsection{Evaluating in between Configurations}
\label{sec:EvalutingInBetweenConfigurations}

The definition and implementation of boundary configurations set our frame of reference in terms of effort and especially data quality. During the evaluation phase, we developed an idea annotation tool which is open source and published at~\url{https://github.com/FUB-HCC/Innovonto-ICV}. Moreover, all data from the studies are also openly available at~\url{https://osf.io/k2ey7/}. We next continue with describing the development of the in between configurations in more detail.

\subsubsection{User Interface}
\label{sec:userInterface}
We built software to annotate concepts via a web interface as a baseline employed for all configurations.
Furthermore, the software enables the use of the configurations previously defined for the tasks \textit{Concept Ranking} and \textit{Concept Selection}, due to the potential benefits of human judgement in \textit{word-sense disambiguation}.
Subsequently, we describe both the prototype and algorithmic realization of the first two steps in the task flow.

\begin{figure}[tb]
	\begin{center}
		\includegraphics[width=\columnwidth]{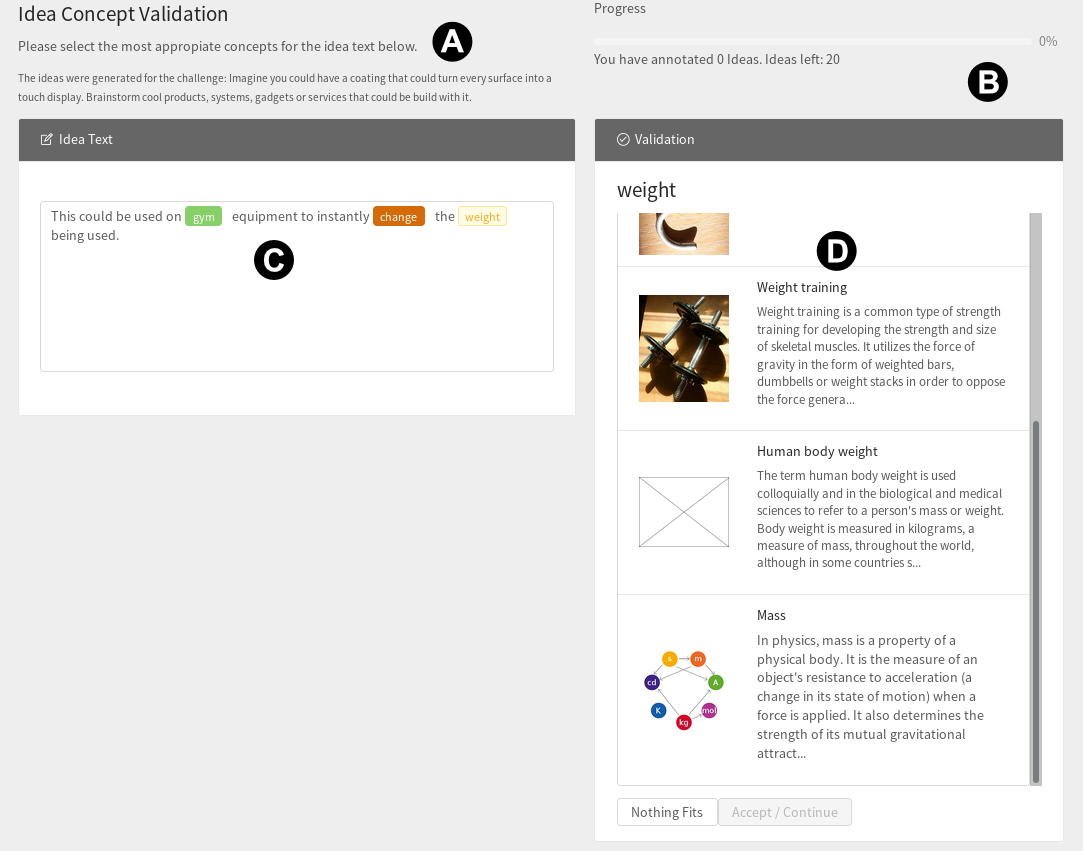}
		\caption{Example Interface for the ICV prototype: Consisting of a task description (Panel \textbf{A}), progress indicator (Panel \textbf{B}) and concept validation interface.
			This example shows an idea text having three possible annotations (`gym,' `change' and `weight').
			The user has already selected a concept for `gym' (as shown by the green background in Panel \textbf{C}) and rejected all concepts for the term `change' (as shown by the orange background.
			Currently, the user has to annotate the term `weight' and, therefore, is shown a list of possible concepts for this term in Panel \textbf{D}.}
		\label{fig:icv-in-action}
	\end{center}
\end{figure}

\subsubsection{Realization of the Term Extraction and Concept Search Tasks}
Figure \ref{fig:icv-in-action} shows the prototype during an active concept annotation task.
The interface consists of four panels (each denoted by a capital letter): Panel \textbf{A} shows the task description and background information about the ideas.
Panel \textbf{B} shows the user's process within the overall task.
Panel \textbf{C} shows highlighted terms in the source text. Terms in the input text are highlighted depending on their current state:
Terms linked to a concept are shown with a green background, terms that are rejected (the user could not find a fitting concept) are shown with an orange background and the currently active term is shown by a yellow background.

Panel \textbf{D} shows the list of possible concepts for the current term.
We employ an automatic search approach in the DBpedia knowledge graph to realize automatic term extraction and concept search\footnote{We used the candidate search API of DBPedia Spotlight (\url{http://api.dbpedia-spotlight.org/en/candidates}) with a confidence threshold of 0.01 to obtain a ranked list of terms and concepts.}.
In the case of overlapping concept term combinations\footnote{Exemplarily, the term `pet food distribution center' could be described by the concepts `pet food,' `food distribution' or `distribution center' (among others).}, we employ a greedy algorithm to expand the term in the idea to the longest continuous text.
This enables multiple annotations of a single term.

The search for concepts results in a link between a term (or set of terms for overlapping results) that is potentially linked to a list of concepts (this is called an \textit{annotation candidate}).
If the search for a term does not result in any concepts, the annotation candidate is discarded.

\subsubsection{Realization of the Concept Selection Task}
We highlight each annotation candidate in the idea text (cf. colored boxes in Panel \textbf{C}) and then ask the user to validate them one by one.
We show the list of concepts found for the currently focused term in Panel \textbf{D}.
For each term, the person has to go through the list of concepts and select the best fitting concepts by clicking on them.
After this step, the person has to submit the concepts selected by clicking~ \texttt{[Accept/Continue]} or, if there is no concept that fits, reject all concepts for the term by clicking~ \texttt{[Nothing\_Fits]}.
After a term is handled, the system automatically selects the next unhandled term.
Once the person has validated all terms, they use the \texttt{[Accept/Continue]} button again to save the annotation results. 

After evaluating different ways to display concepts to the user\footnote{When comparing three different concept representations, we found that image and description led to the best data quality results; for more information, see the technical report at \url{https://refubium.fu-berlin.de/handle/fub188/25266}.}, we chose to represent each concept by its image (if present) and description, obtained from the knowledge graph.
For the concept `Computer Keyboard', for example, we can obtain the following description from DBpedia\footnote{Please check out the concept URL for further information \url{http://dbpedia.org/page/Computer_keyboard}.}:
\textit{`In computing, a computer keyboard is a typewriter-style device which uses an arrangement of buttons or keys to act as a mechanical lever or electronic switch. [$\ldots$].'}

%
%


Based on the in between configurations defined in Section~\ref{sec:applied-loa}, we defined five study conditions that correspond to the different LoA defined:
(1) A baseline condition that corresponds to the LoA 2 state;
(2) a ranked by confidence condition; 
(3) a threshold preselection condition integrating the user; 
(4) a fully automatic threshold selection condition; and 
(5) a combined condition that integrates the ranking and automatic threshold conditions (cf. Table \ref{tab:HM_ICV}).
All conditions are explained in detail next. 

\paragraph{LoA 2 -- Baseline Condition.}

The first configuration was used as a baseline condition and a representative of LoA 2: 
\textit{`The computer offers a complete set of decision/action alternatives'} (cf. Figure \ref{fig:loa_parasuraman}).
Only the tasks \textit{Term Extraction} and \textit{Concept Search} are automated in the baseline system.
The concept candidates obtained from the knowledge graph are shown in alphabetical order (by concept label).

\paragraph{LoA 3 -- Ranking Condition.}
We extended the annotation system from the previous condition by ranking the concept candidates by confidence, i.e., we show concept candidates with a higher confidence score higher up in the list. We interpreted this ranking as a narrowing of the selection (as described in LoA 3), as lower confidence results can only be detected by scrolling down the list.

\paragraph{LoA 4 -- Validated Threshold Annotation.}
We further increased the level of automation by reducing the manual effort in the \textit{Concept Selection} sub-task.
\revision{This condition corresponds to the configuration shown in Figure~\ref{fig:loa}d: The tasks \textit{Concept Search} and \textit{Concept Ranking} are allocated to the computer, and furthermore, the computer preselects a concept for the user. The computer preselected concepts in the implementation with a confidence score higher than  a $0.95$ threshold.}
The user was shown the concepts selected by the computer (indicated by a light green background) and asked to verify them manually, i.e., approve or reject the automated recommendations.
In this condition, we presented the list of concept candidates in alphabetical order as in the \textit{Baseline} condition. 

\paragraph{LoA 5 -- Automatic Threshold-based Annotation.}
The next automation step was to further automate the threshold-based selection of concepts.
In this condition, the computer automatically selected concepts with a confidence value higher than $0.95$.
Furthermore, the automatically annotated term/concept combinations were not shown to the user. Instead, the software skipped over the term and went on to the next unhandled one. 
\revision{This meant that users had to select concept candidates only for terms, when the system had not already selected a concept.
	Users still had the possibility of changing term/concept combinations already annotated by clicking on the term skipped.} 
The concept candidates for the terms not automatically annotated were ordered alphabetically.

\paragraph{LoA 5 -- Ranking and Automatic Threshold Combined.}
The last condition combines the automation features described in the previous two paragraphs to increase the computer contribution further.
In this \textit{combined} condition, we automatically select concept candidates with a confidence threshold higher than $0.95$ and ordered all concept candidates by their confidence scores, respectively. \revision{While this combination of automation features represented a new condition in the systematic evaluation, in terms of abstract LoA, this configuration still corresponds to a LoA of 5.}

\subsubsection{Study Setup}
\label{sec:study-setup}
We conducted a preliminary study where participants received a random selection of 20 ideas from our gold standard corpus containing 60 ideas.
As each participant received different ideas, results for the different conditions showed a very high variance across all metrics.
We assume that the results of the data annotation task were highly dependent on the randomly chosen idea (e.g., some ideas are concise and, therefore, harder to understand).
We, therefore, provided all participants with the same 20 ideas in the same order across all conditions\footnote{See Appendix~\ref{app:ideas} for the idea texts.}.
This decision of focusing on the annotation task prevented us from obtaining ideation metrics as described in Section~\ref{sec:metricsHumanPerspective}. While these metrics are important in the domain context of collaborative ideation, we chose the new task design to ensure comparability between conditions for the study discussed subsequently.


We used Amazon Mechanical Turk (MTurk) to recruit participants.
One of the goals of the system tested was the usage in an microtask/crowdworker environment, so it had to be usable without extensive training.
\revision{We chose to employ a between-group experimental design to prevent learning effects.
	This study design, furthermore enabled us to create HITs\footnote{HIT: Human Intelligence Task - One MTurk task.} comparable to other annotation tasks on MTurk in complexity and time, limiting worker frustration.}
We recruited 40 participants for each condition via MTurk (200 persons in total). \revision{We limited participants to US residents who had completed at least 1,000 HITs with a greater than 95~\% approval rate to ensure high task quality. Participants were paid \$4.5 (\textasciitilde\$12/h)  for a session in one condition.}

The task consisted of a short tutorial on concept validation followed by the annotation of 20 ideas.
We employed the interface shown in Figure \ref{fig:icv-in-action} for all conditions.
After submitting the annotations of an idea, the system moved to the next idea text automatically. After the annotation of the last idea, the system displayed the survey page.
In the survey, the participants were asked to provide selected demographic data (e.g., age group, gender), fill out the questionnaire about their experience and, optionally, provide qualitative feedback.
We used the metrics and questionnaires described in Section \ref{sec:metrics} for all conditions.

\paragraph{Hypotheses}
The hypotheses for this study trace back to our original model of a point of friction between data quality and human effort.
We generally expected this point of friction to show up in metrics as well. 
Regarding click effort, we expected the conditions featuring a threshold mechanism to have a lower rate of recorded clicks when compared to the \textit{Baseline} condition ($H_1$). 
Regarding the \textit{Ranking} condition, we expected a similar outcome to the \textit{Baseline} condition in terms of click effort, as no concepts are automatically annotated by the system.
Regarding time effort we expected a lower time when comparing the \textit{Baseline} and \textit{Ranking} condition, as participants find the most fitting concepts at the top of the list ($H_2$).
Furthermore, we expected a reduced time for the conditions featuring a threshold mechanism as well ($H_3$).
The lowered effort in time and clicks should reflect on the perceived effort as well, so we expected a lower effort in the \textit{Ranking}, \textit{Validated Threshold}, \textit{Automatic Threshold} and \textit{Ranking/Threshold} conditions when compared to the \textit{Baseline} ($H_4$).

Regarding data quality, we expected a lower F-measure for the conditions featuring an automatic threshold mechanism when compared to the \textit{Baseline} and \textit{Ranking} conditions ($H_5$), as the system could have selected concepts incorrectly.
An interesting case is provided by the \textit{Validated Threshold} condition, as participants are able to correct concepts incorrectly selected by the system.
We, therefore, expected the \textit{Validated Threshold} condition to show lower human effort metrics than the \textit{Baseline} and \textit{Ranking} condition, while simultaneously producing higher data quality than the \textit{Automatic Threshold} and \textit{Ranking/Threshold} conditions ($H_6$).

\subsubsection{Results}
\label{sec:study2-results}
The input data set consists of 20 ideas and 152 terms with possible concept annotations.
As we used a between-group study design, these annotations were produced redundantly by 40 participants per condition.
We excluded the following participants from the analysis to ensure the quality of the results and filter our results that are most likely spam:
Participants who rejected all annotations (total of 4), and those who did not reject a single annotation (total of 2) were excluded. The second filter was possible because the gold standard contained at least one concept that was not in the list of candidates during the deployment of the system.
Table~\ref{tab:results-study-3} shows the number of participants included per configuration.
Table~\ref{tab:results-study-3} also shows the metrics obtained for all conditions described previously in terms of the number of clicks, time needed, TLX score, precision, recall and F-measure.
We report clicks, time and TLX score as absolute values as we predefined the number of annotations.

Regarding effort metrics, we see that mean click effort is lower for the higher automation conditions \textit{Validated Threshold}, \textit{Automatic Threshold} and \textit{Ranking/Threshold}.
The lowest mean number of clicks is reported for the \textit{Automatic Threshold} condition.
The lowest average time spend in concept validation is measured in the \textit{Automatic Threshold} condition. 

We conducted an analysis of variance (ANOVA) for click effort between all conditions ($F=16.98, p<0.001$) and then conducted a Tukey HSD to validate $H_1$.
The results of the Tukey HSD show a significant difference in the click effort between the \textit{Baseline} and \textit{Automatic Threshold}, \textit{Validated Threshold} ($p<0.001$) and the \textit{Ranking/Threshold} ($p<0.001$) conditions.
We conducted an ANOVA for time effort between all conditions ($F=2.36,p=0.05$) to validate $H_2$ and $H_3$.
We then conducted a Tukey HSD on the results.
When comparing the \textit{Baseline} condition and \textit{Ranking} condition, we found no significant difference in time effort ($p=0.99$).
When comparing the \textit{Baseline} condition to the automatic threshold conditions regarding time effort, we found significant differences to the \textit{Validated Threshold} condition, the \textit{Automatic Threshold} condition ($p<0.01$) and the \textit{Ranking/Threshold} ($p<0.01$) condition.
We conducted an ANOVA for TLX scores between all conditions.
Results show no significant differences between conditions ($F=0.384,p=0.82$) to validate $H_4$.
We conducted an ANOVA for the F-measure between all conditions to evaluate the impact of the automation features on data quality ($H_5$).
The results show no significant differences between the conditions ($F=1.054, p=0.38$).



\begin{table}[t]
	\footnotesize
	\begin{tabular}{p{1.2cm}rrrrrrrrrr}
		\toprule
		Configs. & \multicolumn{2}{c}{Baseline} & \multicolumn{2}{c}{Ranking} & \multicolumn{2}{c}{\begin{tabular}[c]{@{}c@{}}Validated\\ Threshold\end{tabular}} & \multicolumn{2}{c}{\begin{tabular}[c]{@{}c@{}}Automatic\\ Threshold\end{tabular}} & \multicolumn{2}{c}{\begin{tabular}[c]{@{}c@{}}Ranking/\\ A. Threshold\end{tabular}} \\
		\midrule
		LoA & \multicolumn{2}{c}{2} & \multicolumn{2}{c}{3} & \multicolumn{2}{c}{4} & \multicolumn{2}{c}{5} & \multicolumn{2}{c}{5} \\
		Participants & \multicolumn{2}{c}{40} & \multicolumn{2}{c}{40} & \multicolumn{2}{c}{36} & \multicolumn{2}{c}{39} & \multicolumn{2}{c}{39} \\ 
		&&&&&&&&&& \\
		& \multicolumn{10}{c}{\textit{Human Perspective (Effort)}} \\ \midrule
		Clicks & 314.25 & (67.21) & 304.68 & (51.51) & 284.61 & (53.33) & \cellcolor{gray!25} 230.38 & \cellcolor{gray!25} (45.92) & 229.05 & (77.29) \\
		Time (s) & 1,528.42 & (528.51) & 1,449.78 & (577.02) & 1,379.78 & (600.28) & \cellcolor{gray!25} 1,139.87 & \cellcolor{gray!25} (560.70) & 1,271.08 & (620.44) \\
		TLX & 20.88 & (5.64) & 20.32 & (4.17) & 20.44 & (3.83) & \cellcolor{gray!25} 20.26 & \cellcolor{gray!25} (4.44) & 20.82 & (4.57) \\
		&&&&&&&&&& \\
		& \multicolumn{10}{c}{\textit{Algorithmic Perspective (Data Quality)}} \\ \midrule
		Precision & \cellcolor{gray!25} 0.51 & \cellcolor{gray!25} (0.29) & \cellcolor{gray!25} 0.51 & \cellcolor{gray!25} (0.29) & 0.50 & (0.28) & 0.46 & (0.27) & 0.45 & (0.27) \\
		Recall & 0.52 & (0.26) & 0.52 & (0.26) & 0.53 & (0.26) & 0.51 & (0.27) & \cellcolor{gray!25} 0.54 & \cellcolor{gray!25} (0.26) \\
		F-measure & 0.48 & (0.25) & 0.49 & (0.26) & \cellcolor{gray!25} 0.50 & \cellcolor{gray!25} (0.26) & 0.46 & (0.25) & 0.47 & (0.25) \\ \bottomrule
	\end{tabular}%
	\caption{Evaluation results for Effort and Data Quality metrics for different automation configurations.
		While Effort metrics are lower for higher LoA, mean data quality (F-measure) is highest for the \textit{Validated Threshold} configuration.
		All results are reported in mean ($\pm$ standard deviation).}
	\label{tab:results-study-3}
\end{table}

In addition to the analysis of human effort and data quality, we wanted to get a better understanding of the attitudes the users have towards the system.
Table~\ref{tab:results-resque} shows the results for the ratings received for the questionnaire employed.
All questions use a 5-point Likert scale ranging from -2 (Strongly Disagree) to 2 (Strongly Agree).
\revision{The \textit{Ranking/Threshold} condition was rated higher on average than the other conditions for 6 out of the 11 questions.
	In terms of adequacy of interface layout and expectations, the \textit{Baseline} condition was rated highest on average. Regarding information sufficiency, the \textit{Ranking} condition was rated highest. Regarding the diversity of displayed items, the \textit{Validated Threshold} condition was rated highest. While the \textit{Baseline} condition was rated highest in trust, the \textit{Validated Threshold} condition received the lowest score}.

\begin{table}[t]
	\begin{center}
		\footnotesize
		\centering
		\begin{tabular}{@{}L{3cm}R{1.75cm}R{1.75cm}R{1.75cm}R{1.75cm}R{1.75cm}@{}}
			\toprule
			Condition & \multicolumn{1}{l}{Baseline} & \multicolumn{1}{l}{Ranking} & \multicolumn{1}{l}{\begin{tabular}[c]{@{}l@{}}Validated\\ Threshold\end{tabular}} & \multicolumn{1}{l}{\begin{tabular}[c]{@{}l@{}}Automatic\\ Threshold\end{tabular}} & \multicolumn{1}{l}{\begin{tabular}[c]{@{}c@{}}Ranking/\\ A. Threshold\end{tabular}} \\
			LoA & \multicolumn{1}{c}{2} & \multicolumn{1}{c}{3} & \multicolumn{1}{c}{4} & \multicolumn{1}{c}{5} & \multicolumn{1}{c}{5} \\ \midrule
			$Q_{01}$ (Expectations)    & \cellcolor{gray!25} 0.88 (0.82) & 0.41 (1.02) & 0.33 (1.01) & 0.49 (1.10) & 0.64 (1.04) \\
			$Q_{02}$ (Diversity)       & -0.75 (1.10) & -0.90 (1.02) & \cellcolor{gray!25} -0.50 (1.13) & -0.67 (1.15) &  -0.97 (0.99) \\
			$Q_{03}$ (Layout)          & \cellcolor{gray!25} 0.92 (1.07) & 0.85 (1.22) & 0.44 (1.30) & 0.69 (1.32) & 0.74 (1.14) \\
			$Q_{04}$ (Explanation)     & 0.47 (1.26) & 0.12 (1.12) & 0.00 (1.31) & 0.07 (1.29) & \cellcolor{gray!25}0.66 (1.26) \\
			$Q_{05}$ (Information)     & 0.73 (1.15) & \cellcolor{gray!25}0.95 (1.18) & 0.61 (1.20) & 0.74 (1.14) & 0.79 (1.22) \\
			$Q_{06}$ (Interaction)     & 1.05 (0.88) & 1.02 (1.21) & 0.67 (1.20) & 0.74 (1.14) & \cellcolor{gray!25}1.10 (0.97) \\
			$Q_{07}$ (Familiarization) & 1.28 (0.96) & 1.05 (1.34) & 0.89 (1.37) & 0.95 (1.45) & \cellcolor{gray!25}1.36 (0.90) \\
			$Q_{08}$ (Understanding)   & 0.85 (1.14) & 0.90 (1.18) & 0.75 (1.16) & 0.54 (1.29) & \cellcolor{gray!25}1.05 (1.02) \\
			$Q_{09}$ (Ideal Item)      & 0.65 (0.98) & 0.61 (1.24) & 0.36 (1.13) & 0.44 (1.21) & \cellcolor{gray!25}0.72 (1.12) \\
			$Q_{10}$ (Satisfaction)    & 0.85 (1.00) & 0.83 (1.12) & 0.69 (1.21) & 0.72 (1.23) & \cellcolor{gray!25}0.92 (0.98) \\
			$Q_{11}$ (Trust)           & \cellcolor{gray!25}0.97 (0.86) & 0.85 (1.06) & -0.05 (1.19) & 0.43 (1.19) & 0.84 (1.09) \\ \midrule
			Overall Rating        & 0.70 (0.63) & 0.61 (0.85) & 0.38 (0.68) & 0.47 (0.84) & 0.75 (0.73) \\ \bottomrule
		\end{tabular}
		\caption{Evaluation results for the ResQue questions. All results are reported in mean ($\pm$ standard deviation). All scales had a range from -2 (Strongly Disagree) to 2 (Strongly Agree).}
		\label{tab:results-resque}
	\end{center}
\end{table}

\subsubsection{Discussion}
\label{sec:study-discussion}
\label{sec:discussion-icv}
While the results show that increased automation impacts the interaction effort, the results for perceived effort and data quality were inconclusive.
Subsequently, we discuss our main findings from this study.

\paragraph{The Impact of Threshold-Based Annotation on Human Effort.}
The reduction in the number of clicks and time spent on the conditions containing the confidence threshold mechanism implicate an added value of the increased automation in concept annotation.
Participants in the \textit{Automatic Threshold} condition took an average of 622 seconds fewer to validate the concepts when compared to the \textit{Baseline} condition.
Participants in the \textit{Automatic Threshold} condition had an average annotation time of 47 seconds per idea.
Surprisingly, the perceived task load results do not differ across the conditions.
There are two possible explanations for this effect.
Firstly, the individual variance in the reported score was high:
Individual differences in perceived task load per study participant could outweigh the impact of the different conditions.
This could be reinforced by the fact that participants had no frame of reference for their subjective task load due to the between-group design.
Finally, inconclusive results could be due to the sample size. 

%

\paragraph{Understanding Data Quality Variations.}
\label{sec:minmax-analysis}
When comparing the data quality results obtained via fully automatic extraction (cf. Section \ref{sec:automaticedge}) to the data quality of annotations made by study participants, the average F-measure is higher for all semiautomatic approaches.
This shows that the initial motivation for the interactive concept validation approach (algorithmic systems struggle with \textit{word-sense disambiguation}) still holds.
However, when analyzing individual data quality results, we found that while average data quality was higher, some participant results were considerably worse than automatic annotation quality.
There are two possible explanations for this high variance:
Some of the participants might not have understood the task and goal of the concept validation.
Furthermore, participants might not have understood the idea texts and, therefore, could have had trouble to find the correct meaning of terms in the ideas.


\paragraph{The Relationship Between Task Understanding and Data Quality.}
We linked the average F-measures to answers given in the ResQue questionnaire to further discuss data quality results.
An interesting finding for our research was that while participants in the \textit{Ranking/Threshold} condition rated the explanation and trust in the system higher than participants in the \textit{Validated Threshold} condition, the average F-measure in the \textit{Ranking/Threshold} condition was lower.
This could hint at a side effect introduced by higher LoA:
As the computer takes over without an explanation of the effect, the participant accepts the choices made by the system without challenging it.

\paragraph{Users Attitudes towards the System}
\revision{
	The ResQue results point to a different trend overall than the effort/data quality metrics. While both the \textit{Baseline} and the \textit{Ranking/Threshold} conditions generally received good ratings, the \textit{Validated Threshold} condition performed worse. This could be because in this condition, users were directly exposed to errors of the algorithm by means of incorrectly pre-annotated concepts they had to fix. At the same time, in the \textit{Automatic Threshold} configurations, the users were not exposed to mistakes in ranking and, therefore, had more trust in the system.}

\subsubsection{Limitations}
While the study revealed insights into the impact of different automation configurations on human effort and data quality, it was subject to some limitations. 

We set the evaluation task to a fixed set of 20 idea texts, always shown in the same order.
While this study setup was necessary to enable comparative results and make an informed decision about the `sweet spot,' this fixed data set could introduce biases that limit the generalizability to other data sets.
Another possible limitation regarding data quality results could be the understanding of idea texts by the study participants: Siangliulue et al.~\cite{siangliulue2016ideahound} remark that workers had trouble classifying the relatedness of ideas generated by others.


Future work could tackle both limitations by integrating the concept annotation in a brainstorming session so that participants validate concepts in the ideas they provided themselves.
However, metrics have to be adjusted to account for the different idea length and number of concepts annotated by each user to make results of different configurations comparable.

Although we included an animated example showing the annotation of one idea in the task tutorial, for all conditions, the task and its goals were not explained in detail.
One key improvement in this regard could be an interactive tutorial, that ensures participants understand the goals and features of the system before letting them continue to the next step.
Another way to ensure task understanding would be the inclusion of \textit{quality checks}: 
Automatic comparisons between the gold standard and annotated ideas, inserted at specific intervals in the task, that report the data quality result back to the user immediately.
Furthermore, participants could be influenced by a number of factors that affect their concept selection.
We, thus, need an explicit step to define the dependencies in the relationship of the human---computer system.
We discuss this finding further in Section~\ref{sec:discussion-method}.

One limitation in our use case was the need for a gold standard in order to provide data quality measurements.
This need for a gold standard proved to be a major drawback in the overall method application:
In addition to being effort intensive, the correctness of the gold standard lies in the hands of the creators and, as experienced in the use case, is often subject to many discussions.
Furthermore, we experienced the limitation of a `Boolean' approach to concept extraction during our creation of the gold standard:
We decided to allow multiple correct concepts for one term, as multiple definitions were appropriate (especially for colloquial technical terms, such as `screen').
Having a distance-based quality metric (how far is the `correct' concept away from the annotated one) could be helpful in such instances.
\subsection{Decide on the Sweet Spot-Configuration}

In the context of collaborative ideation, we introduced the interactive concept validation with two goals:
While improving the annotation quality when compared to fully automatic approaches, we did not want to burden the users with too many tasks, because concept validation should not distract from the main task of ideation.
While the results of the data quality suggest that the average data quality was slightly higher in the "Validated Threshold" configuration than in the other configurations, the ResQue results show that people were more confused and rated the system lower than in the more automated configurations (LoA 5).
We assume this confusion impacts the overall ideation process and could result in lowering the ideation metrics. 
Thus, after careful consideration of the application context of collaborative ideation we chose the `Ranking/Threshold' configuration as the most suitable.


\section{Discussion of the Method for Human---Computer Configuration Design}
\label{sec:discussion-method}

In the following, we want to reflect critically on the application of the method proposed.
We report our overall experiences and discuss potential areas of improvement for future applications of the method.

\paragraph{Iterative and Incremental Application}

We modeled the task flow incrementally during the iterative applications of the method.
The model acted as a boundary object for the authors to structure the discussions of the various configurations of applying the LoA.
The model allowed us to look at the automation issue from different perspectives (human and computer), which helped us narrow down existing design choices.
Even though all authors are familiar with the task of modeling in the context of software development, the task flow in combination with the LoA improved our understanding of the potential application areas for automation.
In our previous work, we somewhat randomly followed one automation approach in a specific application context.
The method proposed allowed us to decide about each configuration systematically.
In summary, applying our proposed method case turned out to be a valuable approach from both a system designer and developer's perspective.

\paragraph{Documenting Practice}

We defined a specific set of metrics in our use case.
This choice was inherently important for the evaluation process since we decided which characteristics of the human---computer configuration we should capture.
The careful specification of the metrics is, thus, a critical step in the method application.
More guidance is probably needed here, especially in cases, where system designers are less familiar with the application domain.
We can imagine a metrics catalogue which contains known metrics from previous research, their perspective (human or computer) and an example of the area of application.
Such a catalogue could extend existing documentation approaches for data~ \cite{Holland:2018ux,Gebru:2018wh,Bender:2018tr} or learning models~\cite{Mitchell:2019in}.
Bender and Friedman, for example, propose the concept of data statements, which describes a natural language based documentation practice that represents a data set by considering, among others, the origin of the data set, experimental results, and possible bias~\cite{Bender:2018tr}.
Our configuration catalogue could extend such a data statement by the properties of the specific application context.
Furthermore, while using the metrics, we found it challenging to argue about possible influences on the configurations.
In the case of data quality, the results were quite similar.
It was ambiguous, for example, to factor in the task understanding, task motivation and personal differences in performance on the human side, and the quality of the source data, algorithmic performance in concept search and algorithm parameters on the computer side.
We suggest extending the method by a dependency graph for metric definition to make these dependencies more explicit.
This extension could help the team to build a shared understanding of dependencies, and their influences on the metrics.
Such a dependency graph could even be included in the configuration catalogue.
While selecting one metric, possible dependencies could be shown to the system designer.

%
\paragraph{Determining the Sweet Spot}
All in all, applying the method proposed helped us to further understand the design space of human---computer integration in the context of information extraction.
By implementing and evaluating different automation configurations, we gained an understanding about the design choices needed and the possible impact on the overall outcome. The data obtained during the test phase gave us a comparable understanding of system performance in our use case. Nevertheless, deciding on a `sweet spot' was not an easy task.
We held multiple group sessions to determine the `sweet spot' where we used the metrics obtained to discuss the trade-offs involved. The input in these group sessions were, on the one hand, visualizations of the metrics within a configuration (cf. Appendix \ref{app:resque-boxplot}) and, on the other hand, the result overview tables, containing average metrics over all configurations (as reported in Figure \ref{tab:results-study-3}). These inputs provided the starting point for the subsequent discussion of all metrics and their dependencies which led to the final decision.
Choosing the `sweet spot' is highly domain-dependent and should be a qualitative decision based on the insights gained during the process.

\paragraph{Capturing Qualitative Shifts}

The goal of applying the method proposed was to design possible configurations of LoA.
Each configuration represents a hypothesis about the interdependence between human and computer~\cite{Dekker:2002fo}. As already stated, Latour argues that the delegation of formerly human tasks to a computer can change social practices and responsibilities~\cite{Latour:1988co}.
Instead of using a human perspective on the system alone, a field study in which the different configurations are employed might be more suitable for capturing such changes and understanding existing interdependencies. 
Our empirical evaluation is only one possible piece of evidence, but it cannot be conclusive, since it represents only one point in time.

We, therefore, envisage that we can accompany such interactive intelligent systems over their lifetime. This monitoring could be done in an ethnographic study and by indicators that monitor a system state based on the predefined metrics. Such indicators might help to define thresholds at which the different configurations should presumably be used. A deviation from the expected metrics range, for example, could be an indicator for a change in social practice. However, we doubt that a purely data-driven perspective of a system's usage captures such a shift.
All metrics applied in the use case are based on quantitative measurements, even though they describe more subtle issues, as in the case of the ResQue questionnaire~\cite{pu2011user}.
Replacing or extending a human responsibility by an algorithmic component cannot be captured by quantitative metrics alone since such a replacement/extension causes a qualitative change in the human---computer configuration~\cite{Dekker:2002fo}.
Such a qualitative perspective is especially important if we include trust in our evaluative criteria~\cite{deVisser:2018ju}.
Moreover, we imagine that our approach might be included in existing methods such as value-sensitive design (e.g.,~\cite{Friedman:1996gc, Friedman:2017eu}).
The design of human---computer configurations should follow an \textit{action---reflection model} where stakeholder and system designer co-design the system that evolves during the reflection process~\cite{Yoo:2013us}.
The Computer Supported Cooperative Work (CSCW) community has recently started to embrace this work (e.g.,~\cite{Zhu:2018jl}) and more research is needed to understand how different human---machine configuration influences human well-being qualitatively.

\paragraph{Reflecting on Model-Applicability}
The proposed model is largely motivated by a recent proposal of a concept by Farooq and Grudin~\cite{Farooq:2016ey} on human-computer integration. While they argue that more integrative design and evaluation approaches are needed, the article misses providing clear guidance on this. We transport this general idea of human---computer integration to the area of the interactive intelligence systems and here on AI-based software systems. The broad applicability of the model is, thus, limited to systems that are similar to the one investigated in our study. We assume that our method is especially applicable to systems that are similar to interactive recommender systems, interactive and human-centered machine learning, interactive natural language processing, interactive visualizations, etc. In all these systems, the traditional interactive perspective on software design is augmented by algorithmic `intelligence' and vice versa.

While we applied the proposed method to our use case, several issues emerged during the application. Both the model and the LoA are described on a high level of abstraction. The latter enables the applicability to different problem domains and various forms of interactive intelligent systems (cf. examples above). However, while this abstract formulation of the automation level ensures the general applicability of LoA in our context, the concretization of our use case was a source for discussions. These discussions were finally productive, since they helped to delineate the design space further.
This illustrates that while the model shows some ambiguity at this stage, this ambiguity is necessary to have a broader applicability and can be resolved for the specific application domain.


\paragraph{Taking Human---Computer Collaboration beyond Individuals}
The proposed method for the design of human---computer configurations might evoke some broader discussion on its role in CSCW research. The CSCW community focuses on collaboration that can involve a few individuals, groups, organizations and globally distributed online communities. In this context, technology is frequently seen as a supporting instrument that enables collaboration. In human---computer collaboration, the human and the computer are partners~\cite{Jameson:2011fy, Farooq:2016ey}. From this perspective, individuals, groups, organizations or online communities are not entirely human. In the area of online communities, for example, Geiger and Halfaker examined such partnerships of bots and editors in Wikipedia but also the challenges when mixing human and algorithmic governance (e.g.,~\cite{Geiger:2017:OCC:3171581.3134684, Geiger:2010:WSO:1718918.1718941, HalfakerR12}). These considerations might lead to several interesting questions for the CSCW community, such as the degree of artificial partnership needed and wanted (e.g.,~\cite{walsh:2016tu}), the design of the dialogue between a human and a computer (e.g.,~\cite{Liang:2019}), the adaptivity of artificial partners depending on the current needs of the group or community (e.g.,~\cite{Kamar:2012vb, Grudin:2019}), and the algorithmic governance needed in such human---computer collaborations (e.g.,~\cite{danaher:2016hy}). The proposed method for the design of human---computer configurations might help to investigate these collaborative contexts more precisely, where humans and computers, i.e., machines, collaborate to achieve a shared goal~\cite{Horvitz:1999}.

\section{Conclusion}
Defining a way to combine humans and computers effectively has received a lot of attention in research communities of robotics, ergonomics and human factors.
However, related work provides little guidance on how we should design interactive intelligent systems, especially regarding the level of interaction vs. integration.
Based on related work, we proposed a method for defining and evaluating different LoA in interactive intelligent systems.
We applied the method to the task of information extraction in collaborative ideation.
We were able to reach a conclusion about the use of automation in our problem domain by conducting experiments for the defined configurations.
We provide the tool developed and configurations as open source software to enable replication and further adaption. 
Furthermore, we discussed the application of the method and insights gained from the use case.
In addition to highlighting the benefits of the method, we propose three potential enhancements:
The inclusion of a metrics catalogue and indicators to monitor thresholds for system behavior, the introduction of a dependency graph for metric definition, and potential development of the method towards an \textit{action---reflection model}.
\noindent
We believe that the method gives system designers an approach to elaborate the continuum of interactive intelligent systems systematically and to choose the appropriate configuration based on the structured evaluation of configurations.
Admittedly, this method tackles one specific issue in the design of interactive intelligent systems only.
However, it enabled us to make more informed decisions concerning the automation features of our information extraction tool.
We release the ICV tool with a configuration panel, on which the `sweet spot' can be selected based on the application context.
Based on our experience, there is a need for the method proposed in other research areas, such as data visualization and machine learning.
We hope our contribution has provided a first step towards a design framework for the systematic evaluation of implementing human---computer integration.

\begin{acks}
	
	This work is supported by the German Federal Ministry of Education and Research, grant 03IO1617 (`Ideas to Market'). We would like to thank Dr. Abderrahmane Khiat for his input on the Interactive Concept Validation and Aaron Winter for all the work he did to refine and polish the graphical representation of our research.
\end{acks}

\bibliographystyle{ACM-Reference-Format}
\bibliography{CSCW19-ICV-Tool}

\newpage
\appendix

\section{Appendix}

\subsection{Ideas Selected for the Annotation Task (Section~\ref{sec:study-setup})}
\label{app:ideas}
In the following, we list all ideas selected for the annotation task which we obtained from previous studies~\cite{mackeprang2018innovonto}. The idea text was not edited, except for apparent typing errors that could have impacted the concept candidate search.

\begin{enumerate}
	\item This could be used on gym equipment to instantly change the weight being used.
	\item I would have a touchsreen on my wardrobe door that helps me co-ordinate outfits and different outfits for daily wear and different occasions.
	\item You could put it on your car windows to make them darker or lighter for privacy, that would also help with sun glare, you could touch the spot where the sun is shining in your eyes.
	\item I would place the coating on a boxing heavy bag. It would measure the speed of the punch and power of it as well. It would count the number of punches as well. It would give feedback in order to perfect your technique.
	\item Use the coating on medicine bottles to have all the information about the medication and dosing.
	\item It can be sprayed on roads where people speed . That way when they rollover that part of the road it will give them a feedback of how fast they are going and to drive slower
	\item You could put it on a car for businesses or self employed people to advertise. A realtor could add a house address or a hairdresser or restaurant could put a daily special or coupon code.
	\item Coat some on a computer and it will tell you the specs of the machine like its graphics, cpu and ram
	\item Coat glasses with it - can darken to become sunglasses, add messages to be displayed to personalize.
	\item a touch screen on the dryer to see if the clothes are done before the buzzer goes off
	\item A touchscreen plate that could turn a heating element on and off so your food doesn't get cold
	\item A bathroom mirror that allows you to scroll through the news, see the weather, and make or see appointments for that day.
	\item Secondary computer or video monitors can easily be created to cast digital media onto secondary work or viewing surfaces.
	\item Being able to have entertainment while showering or bathing to do anything like watching shows or playing games.
	\item place the material on cups and it will tell exactly what ingredients are in it and calories. it will also tell the person the temperature of the fluid and if the liquid contains any sort of drugs or poison.
	\item replacement windows that utilize touch display, allows kids to write notes or decorate for the holidays
	\item I would use a touch screen on my bed so that it could set an appropriate temperature for my mattress based on the real weather.
	\item Could be used as a key when locked out of your car by putting it on the hood and  entering a password or scanning a fingerprint.
	\item I would use it on my hand to track things like my heart rate or other information that would be useful for exercise, like how far I've walked.
	\item A shirt that can change colors upon using its touchscreen options on the side. It can used to change colors, and fabric type.
\end{enumerate}

\noindent

\subsection{Questions of the Adapted ResQue Questionnaire} 
\label{app:resque}
We list here all questions we used for our adapted ResQue questionnaire. Besides the questions, we provide their topic.


\begin{table}[tbh]
	\begin{tabularx}{\linewidth}{llX}
		\toprule
		& Topic & Question Text \\
		\midrule
		$Q_1$    & Expectations &  The items recommended to me match my expectations. \\
		$Q_2$    & Diversity &  The items recommended to me are diverse. \\
		$Q_3$    & Layout &  The layout and labels of the recommender interface are adequate. \\
		$Q_4$    & Explanation &  The recommender explains why the items are recommended to me. \\
		$Q_5$    & Information & The information provided for the recommended items is sufficient for me to make a selection. \\
		$Q_6$    & Interaction &  I found it easy to tell the system which items are most suitable. \\
		$Q_7$    & Familiarization &  I became familiar with the recommender system very quickly. \\
		$Q_8$    & Understanding &  I understood why the items were recommended to me. \\
		$Q_9$    & Idea Item &  The recommender helped me find the ideal item. \\
		$Q_{10}$ & Satisfaction & Overall I am satisfied with the recommender. \\
		$Q_{11}$ & Trust & The recommender can be trusted. \\
		\bottomrule
	\end{tabularx}
	\caption{Question Texts together with their identifier and topic.}
	\label{tab:ResQueQuestion}
\end{table}
\noindent
Furthermore, the ResQue questionnaire allows for grouping the different topics under specific dimensions: `User Perceived Qualities,' `User Beliefes,' `User Attitudes' and `Behavioral Intentions.' We did not use the `Behavioral Intentions' but listed them for completeness.

\begin{table}[tbh]
	\begin{tabular}{ll}
		\toprule
		Dimension & Question Topic \\ \midrule
		\multirow{5}{*}{User Perceived Qualities} & Diversity \\
		& Layout \\
		& Explanation \\
		& Information \\
		& Familiarization \\ \hline
		\multirow{4}{*}{User Beliefs} & Expectations \\
		& Interaction \\
		& Understanding \\
		& Ideal Item \\ \hline
		\multirow{2}{*}{User Attitudes} & Satisfaction \\
		& Trust \\ \hline
		\multirow{2}{*}{\sout{Behavioral Intentions}} & \sout{Purchase Intention} \\
		& \sout{Use Intentions} \\ \bottomrule
	\end{tabular}
	\caption{The four dimensions of the different question topics with the unused dimension and questions (striked out).}
	\label{tab:ResQueGroups}
\end{table}

\newpage
\subsection{Boxplots for the Adapted ResQue Questionnaire}

\begin{figure}[htb]
	\centering
	\begin{subfigure}{0.25\textwidth}
		\includegraphics[width=\linewidth]{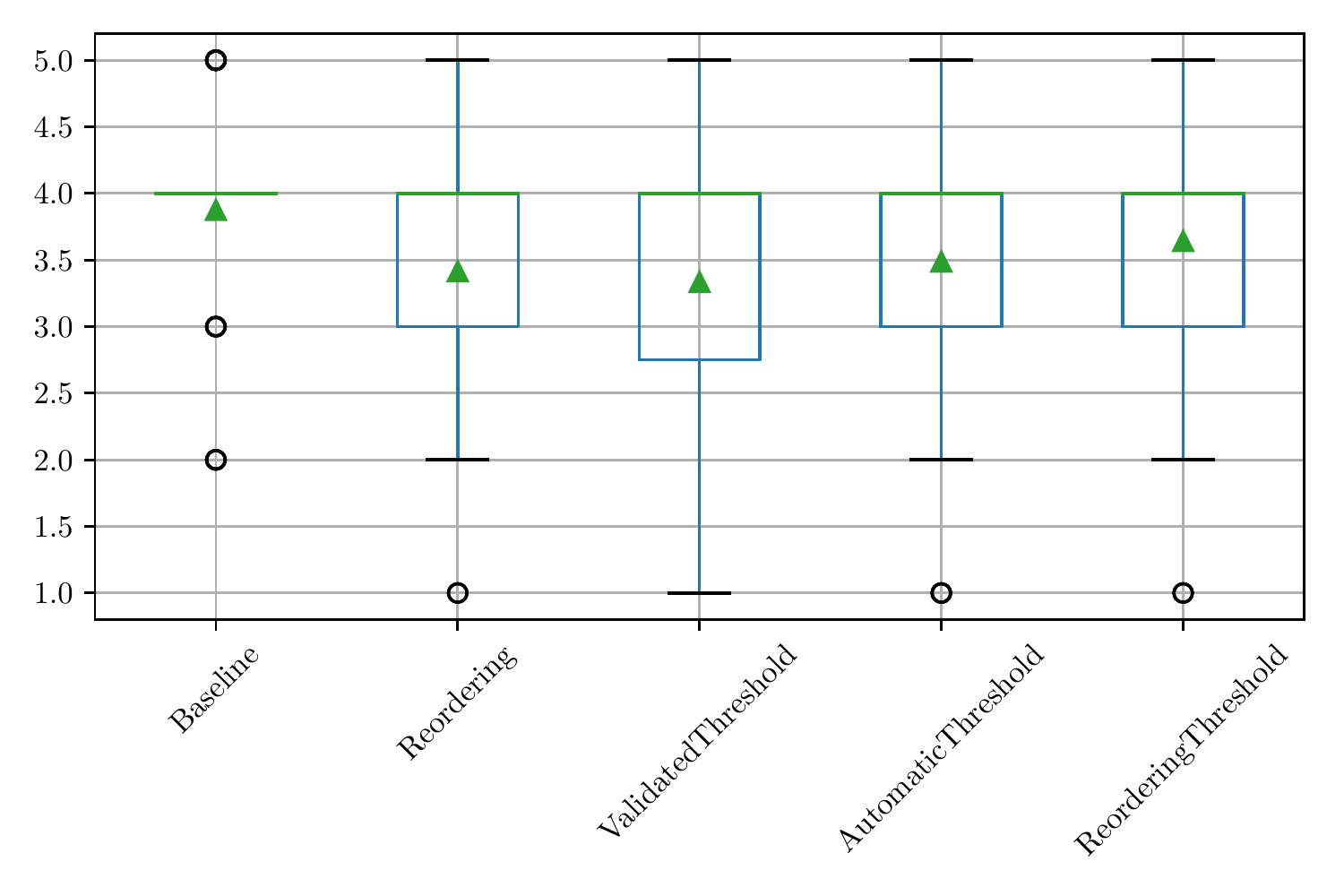}
		\caption{Expectation}
		\label{fig:bp1}
	\end{subfigure}\hfil
	\begin{subfigure}{0.25\textwidth}
		\includegraphics[width=\linewidth]{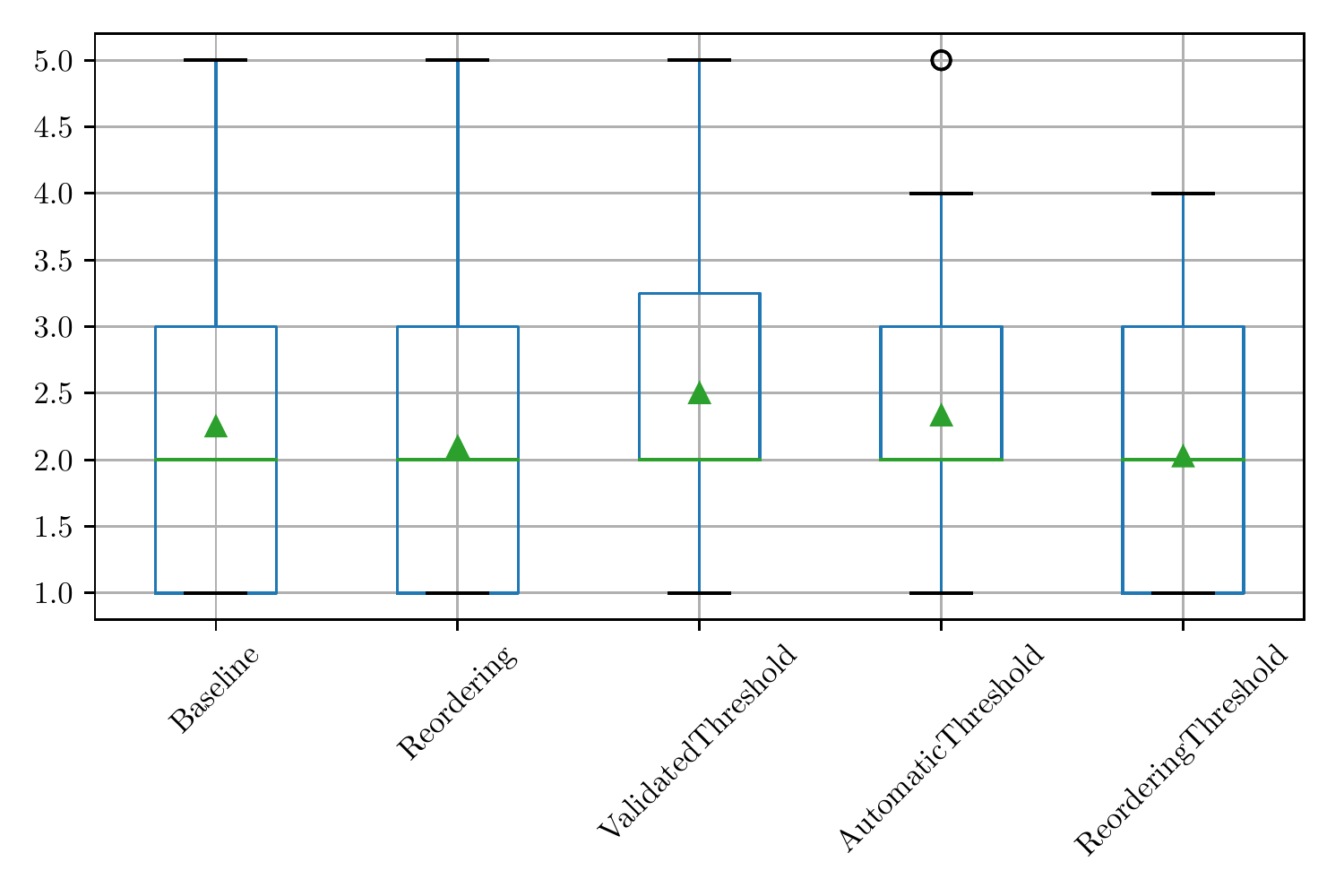}
		\caption{Diversity}
		\label{fig:bp2}
	\end{subfigure}\hfil
	\begin{subfigure}{0.25\textwidth}
		\includegraphics[width=\linewidth]{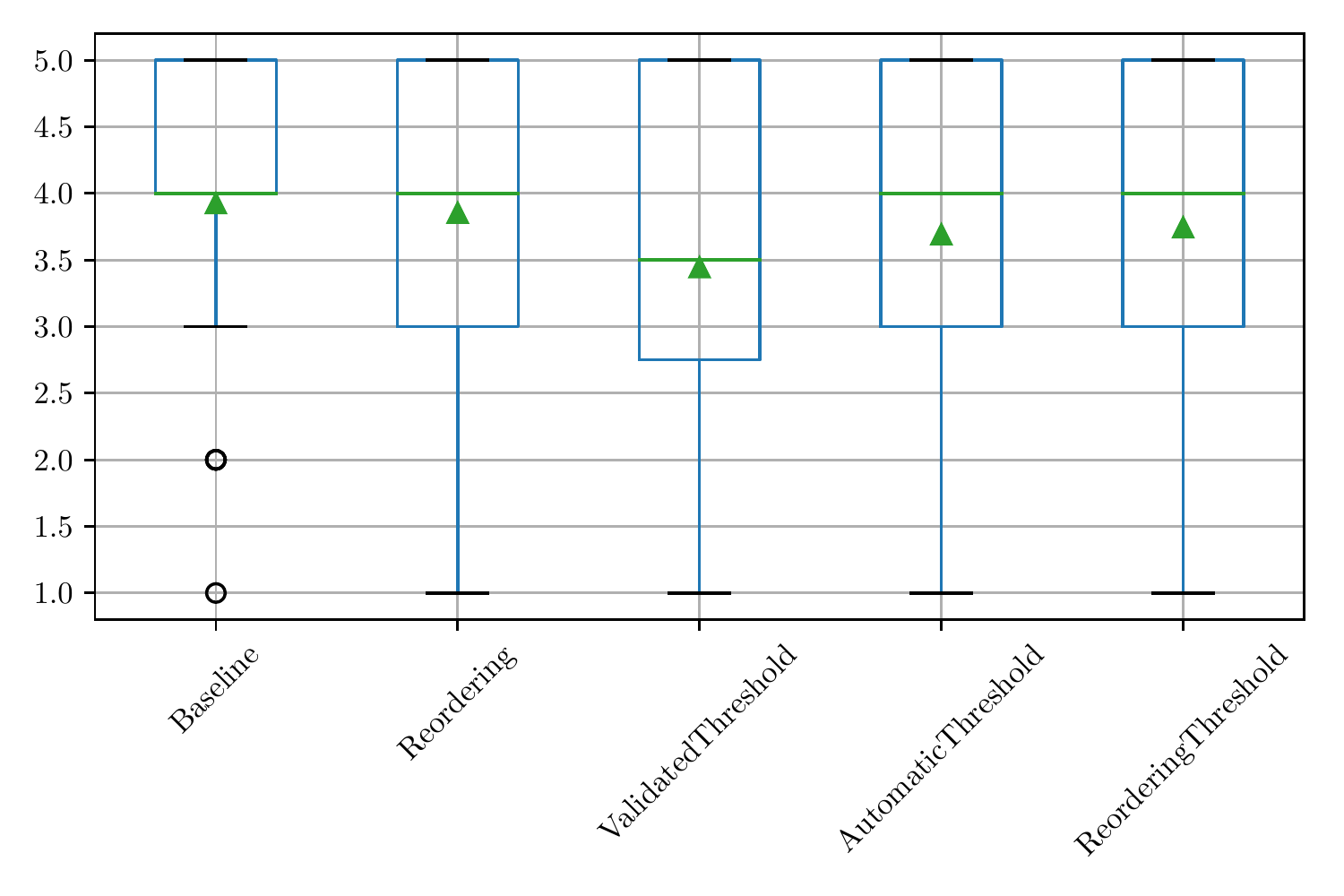}
		\caption{Layout \& Interface}
		\label{fig:bp3}
	\end{subfigure}
	\medskip
	\begin{subfigure}{0.25\textwidth}
		\includegraphics[width=\linewidth,]{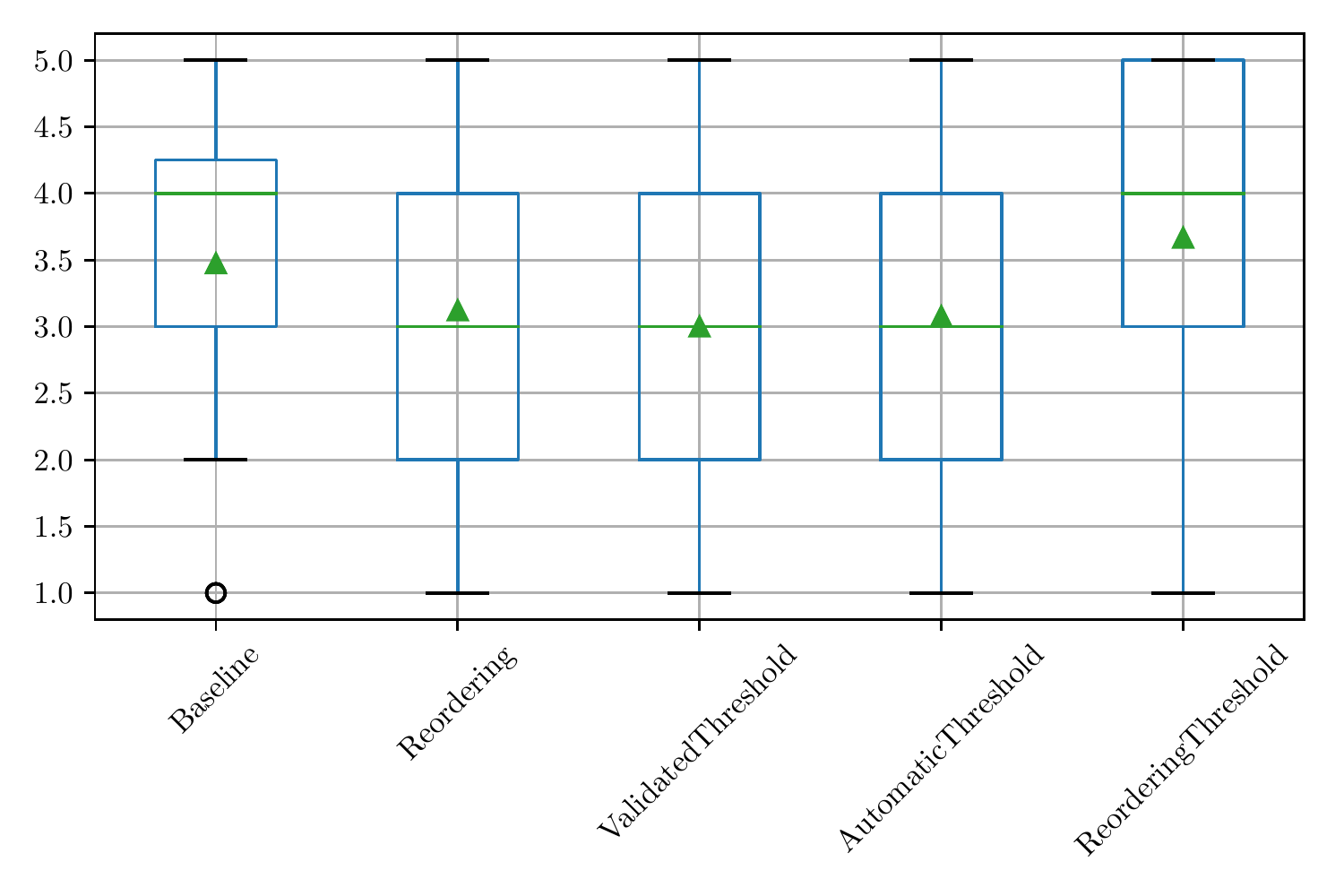}
		\caption{Explanation}
		\label{fig:bp4}
	\end{subfigure}\hfil
	\begin{subfigure}{0.25\textwidth}
		\includegraphics[width=\linewidth,]{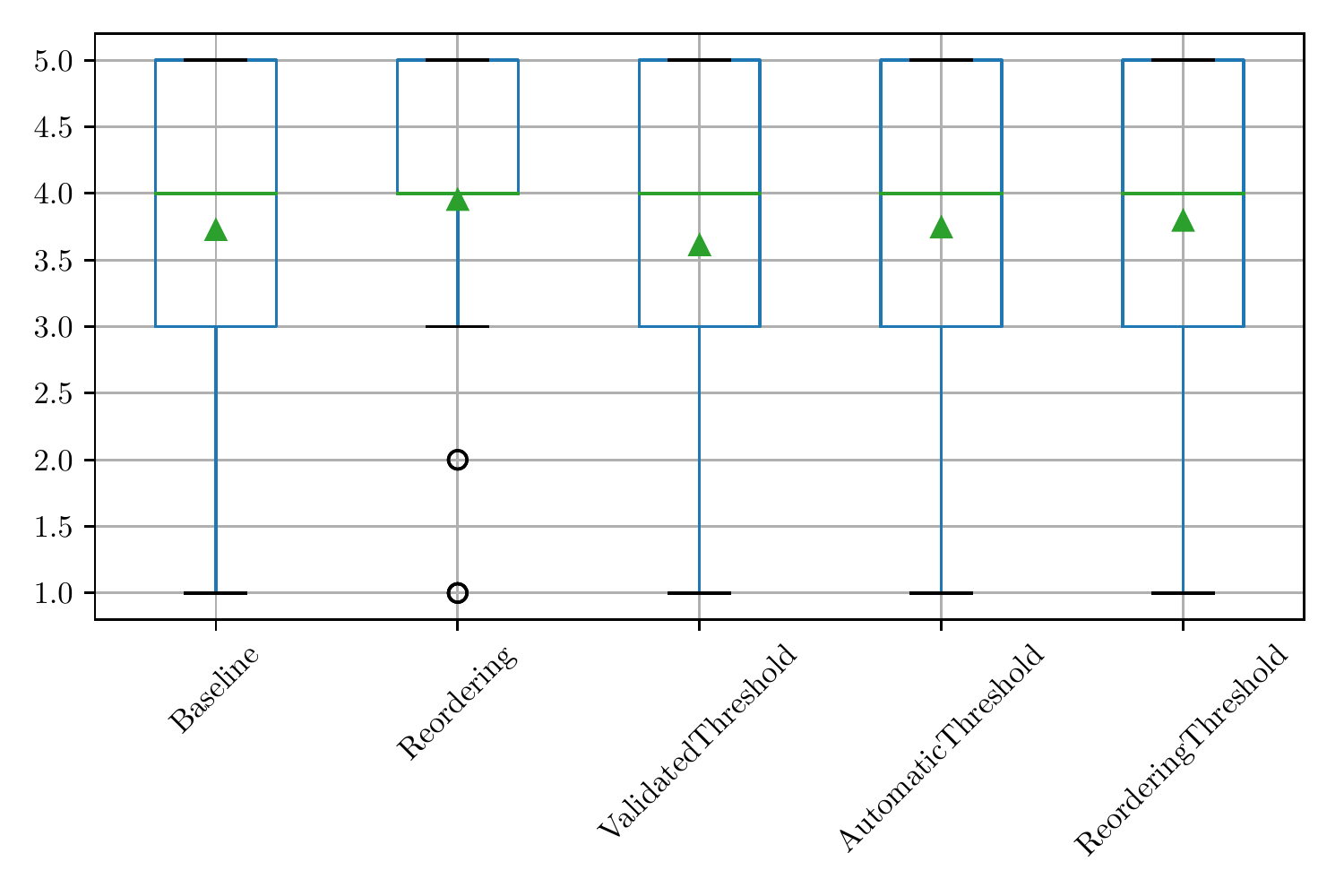}
		\caption{Information Sufficient}
		\label{fig:bp5}
	\end{subfigure}\hfil
	\begin{subfigure}{0.25\textwidth}
		\includegraphics[width=\linewidth,]{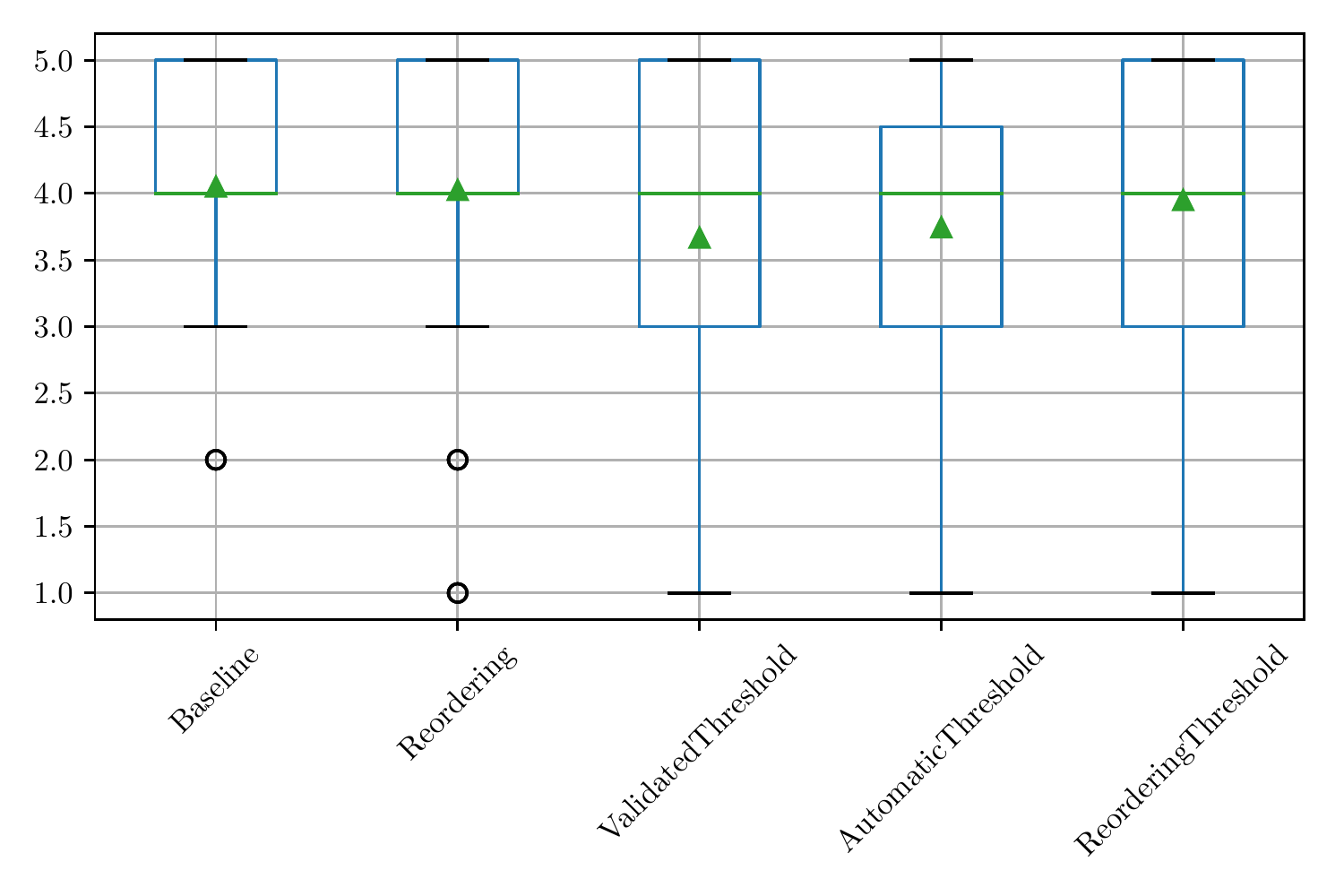}
		\caption{Easy Interaction}
		\label{fig:bp6}
	\end{subfigure}
	\medskip
	\begin{subfigure}{0.25\textwidth}
		\includegraphics[width=\linewidth,]{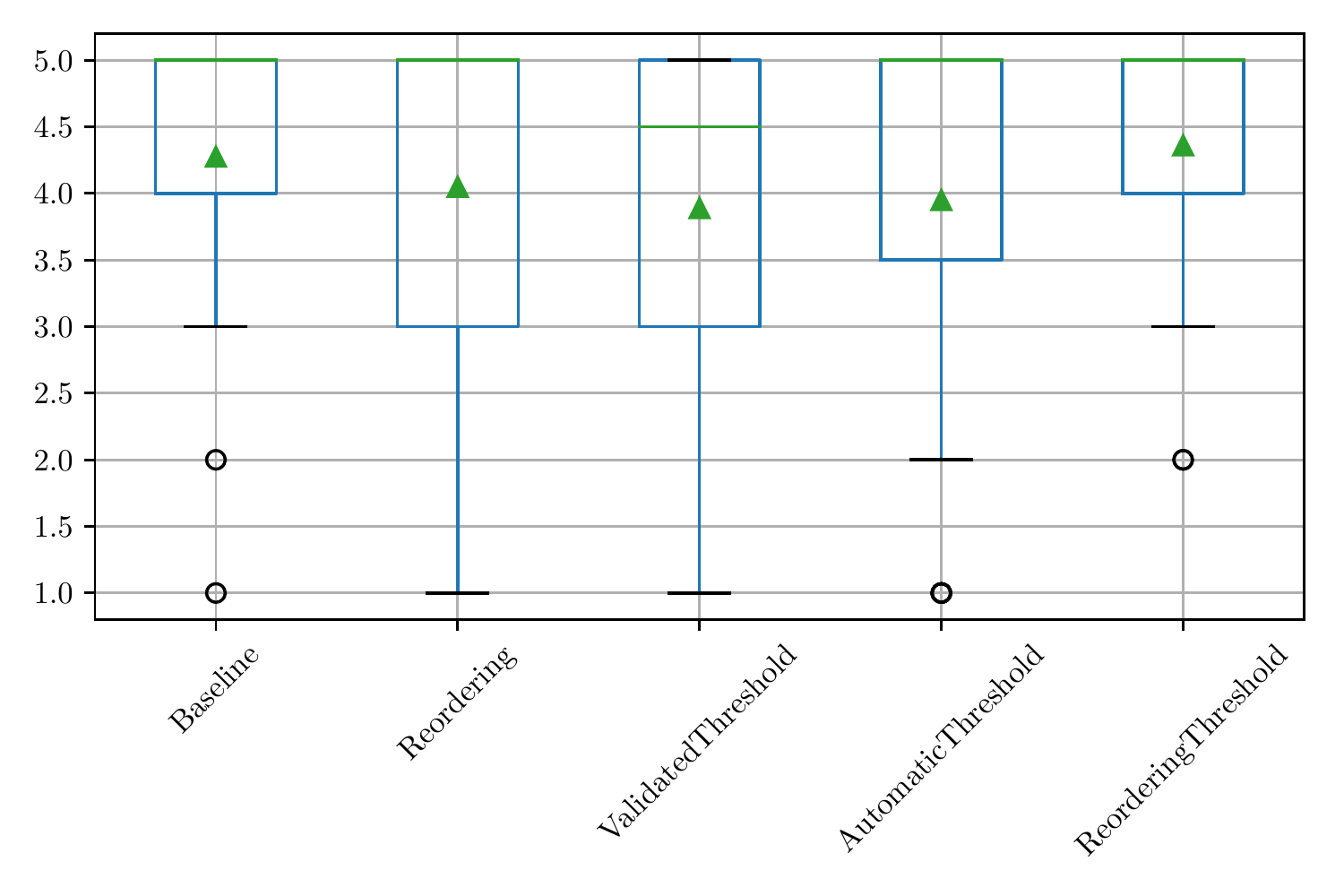}
		\caption{Familiarization}
		\label{fig:bp7}
	\end{subfigure}\hfil
	\begin{subfigure}{0.25\textwidth}
		\includegraphics[width=\linewidth,]{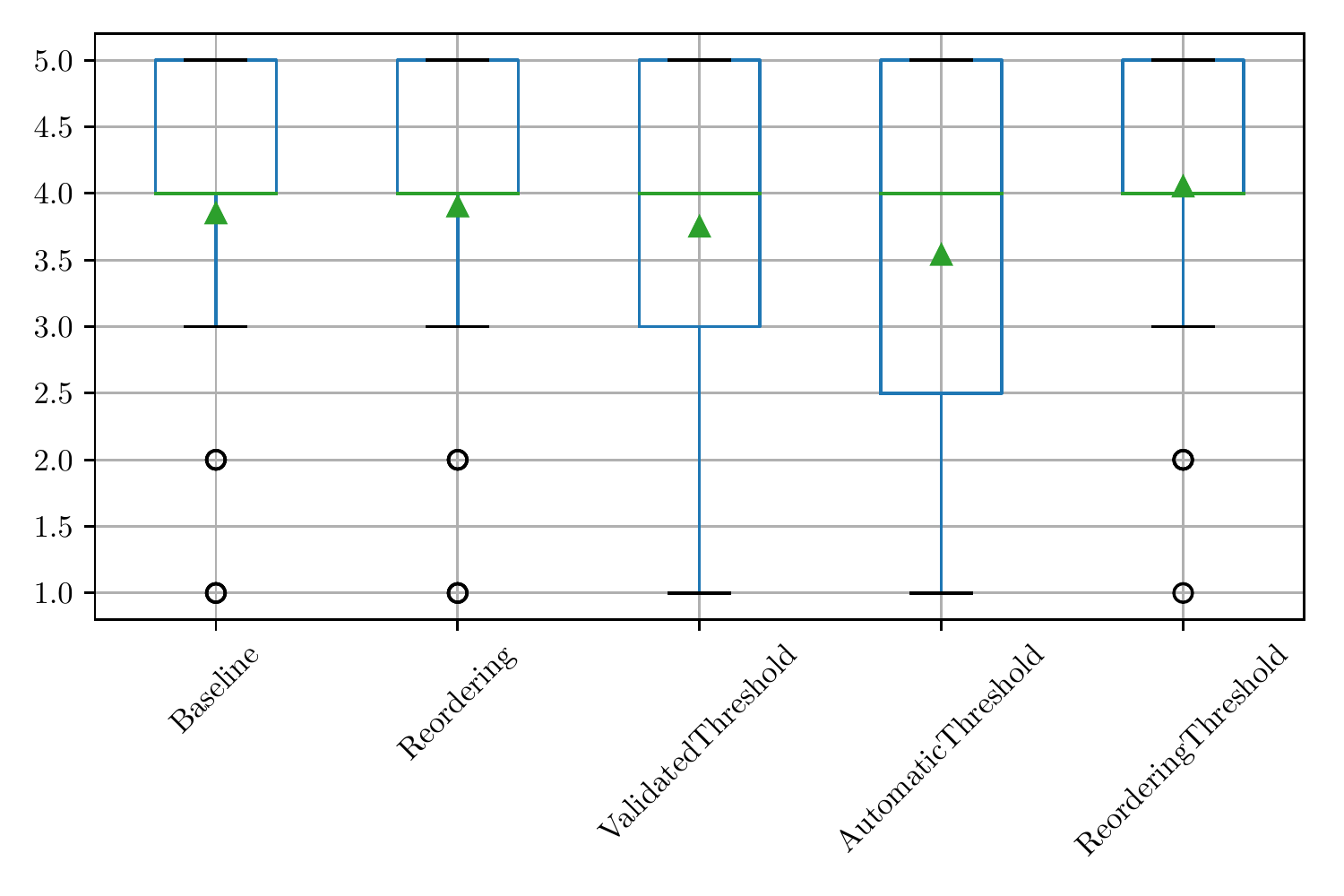}
		\caption{Understanding}
		\label{fig:bp8}
	\end{subfigure}\hfil
	\begin{subfigure}{0.25\textwidth}
		\includegraphics[width=\linewidth,]{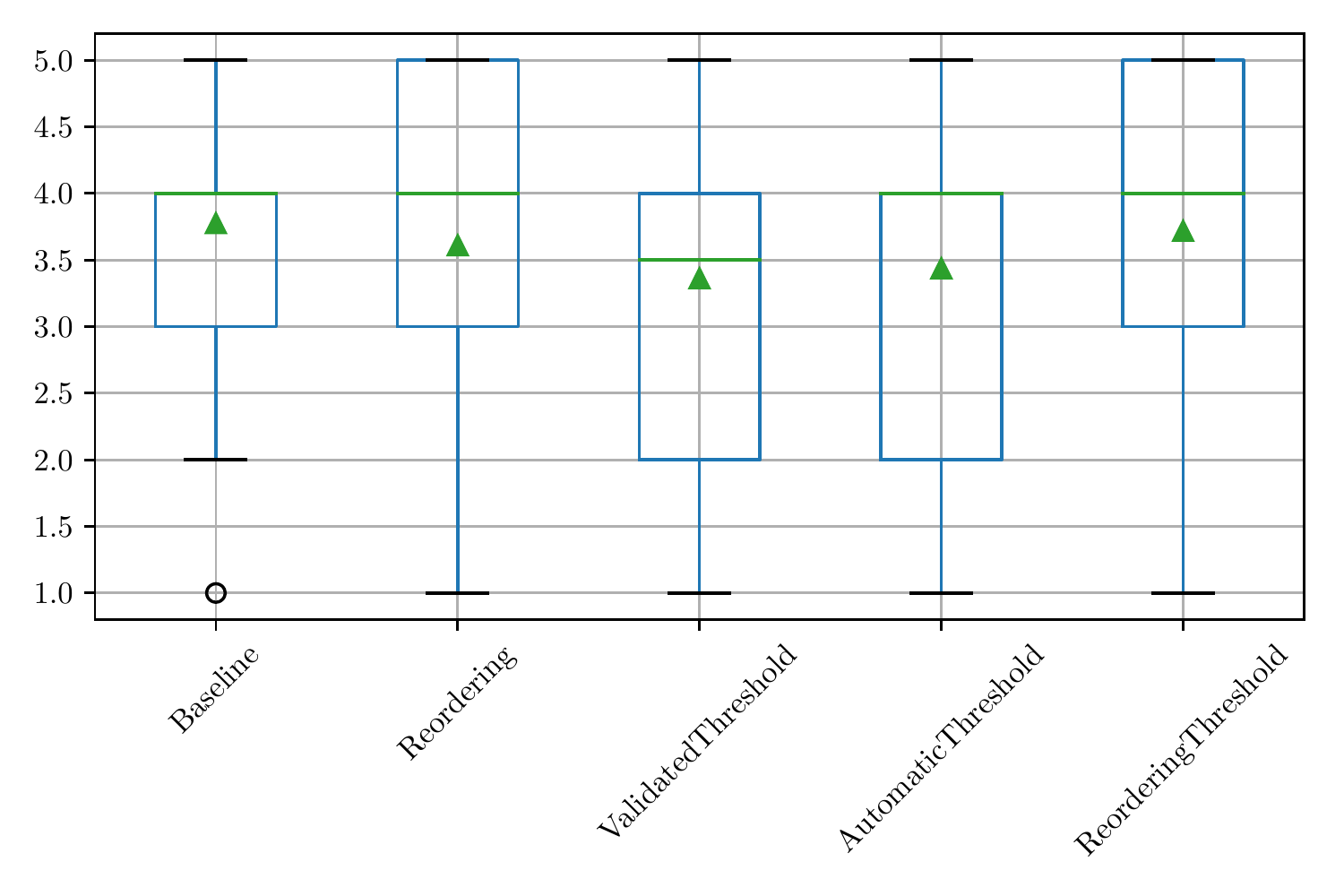}
		\caption{Ideal Item}
		\label{fig:bp9}
	\end{subfigure}
	\medskip
	\begin{subfigure}{0.25\textwidth}
		\includegraphics[width=\linewidth,]{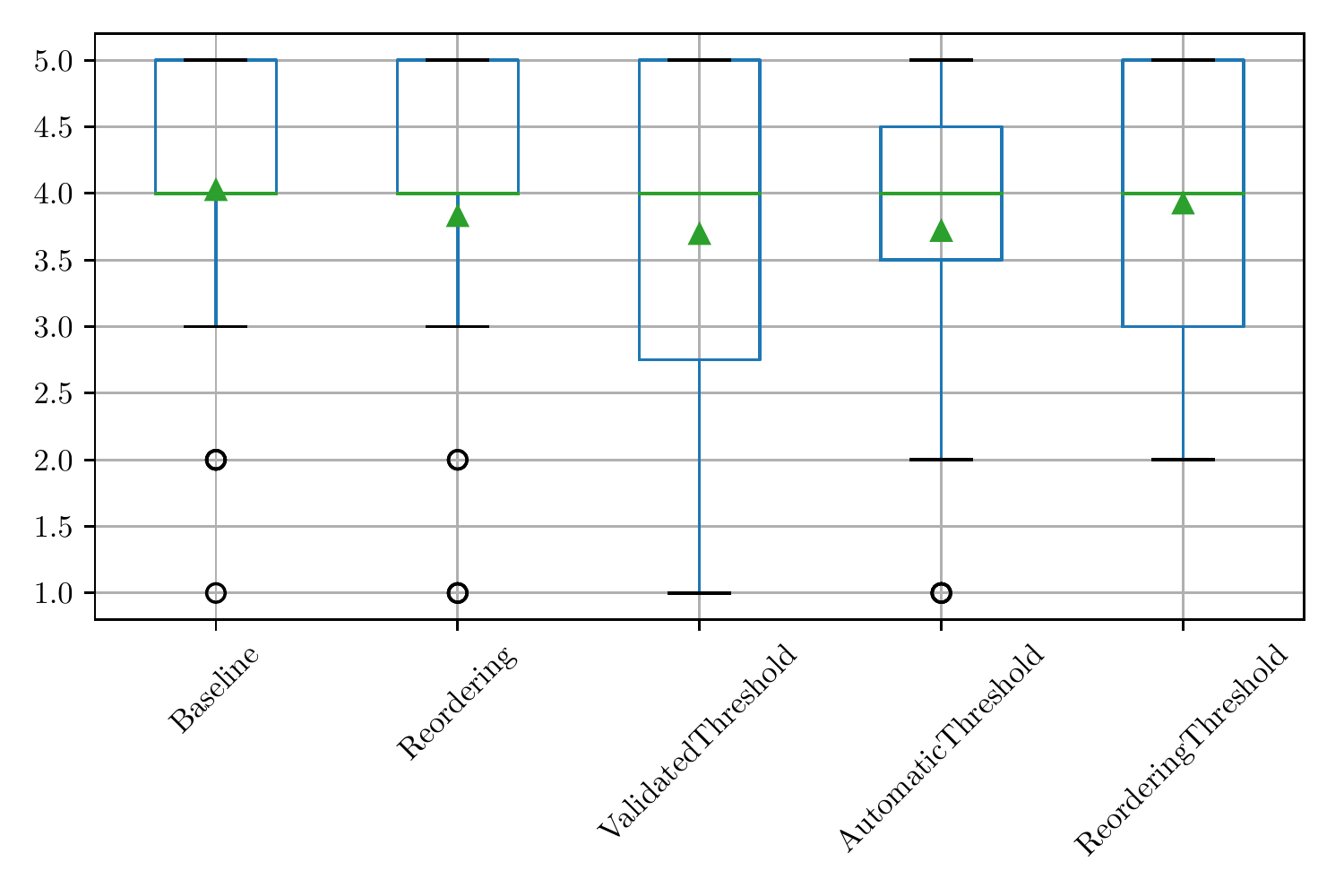}
		\caption{Satisfaction}
		\label{fig:bp10}
	\end{subfigure}\hfil
	\begin{subfigure}{0.25\textwidth}
		\includegraphics[width=\linewidth,]{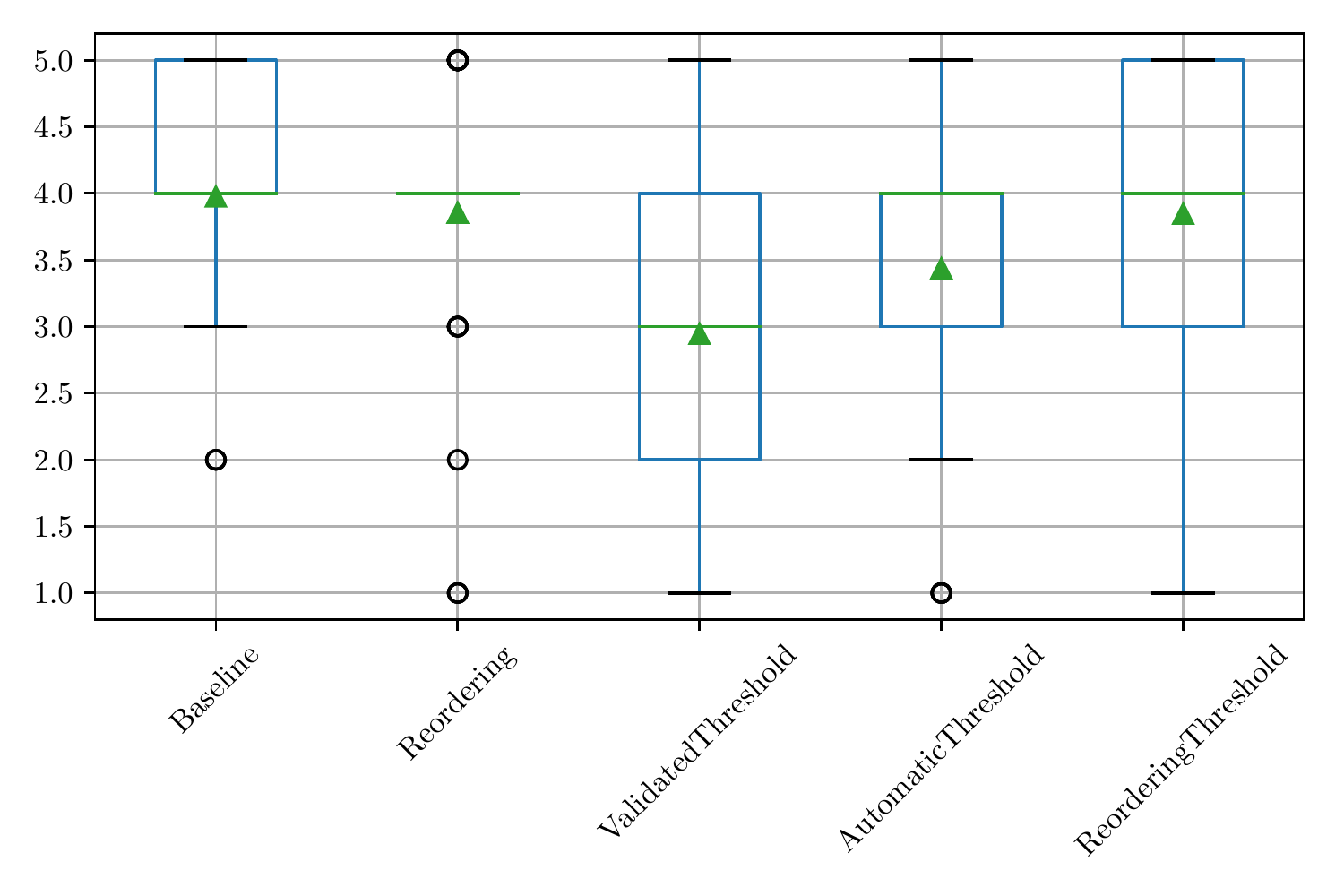}
		\caption{Trust}
		\label{fig:bp11}
	\end{subfigure}\hfil
	\caption{Boxplots for the Adapted ResQue Questionnaire}
	\label{fig:resque-boxplot}
\end{figure}

Figure~\ref{fig:resque-boxplot} shows the boxplots generated for every question topic.

\label{app:resque-boxplot}

\begin{figure}[htb]
	\centering
	\begin{subfigure}{0.25\textwidth}
		\includegraphics[width=\linewidth]{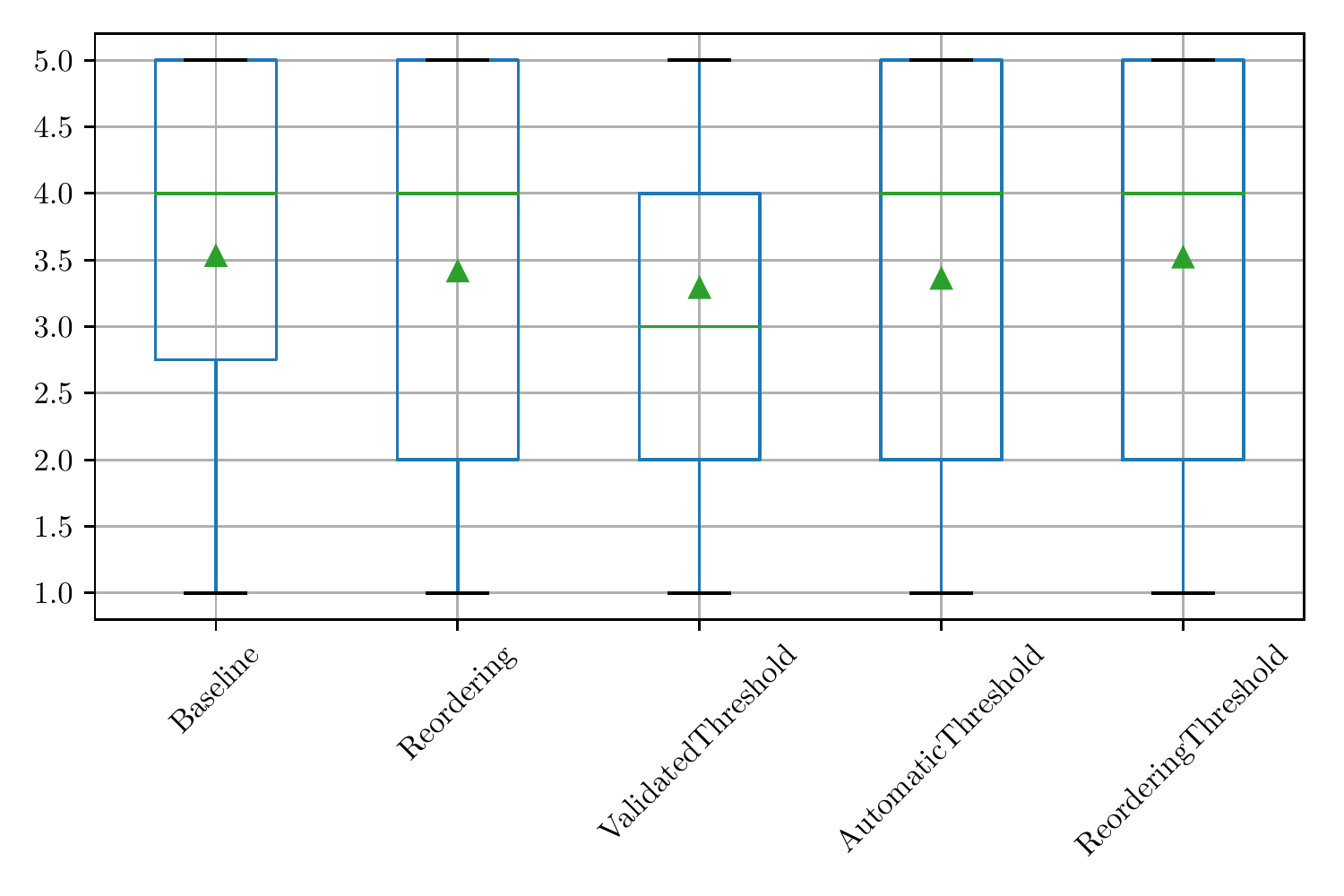}
		\caption{User Perceived Qualities}
		\label{fig:bpg1}
	\end{subfigure}\hfil
	\begin{subfigure}{0.25\textwidth}
		\includegraphics[width=\linewidth]{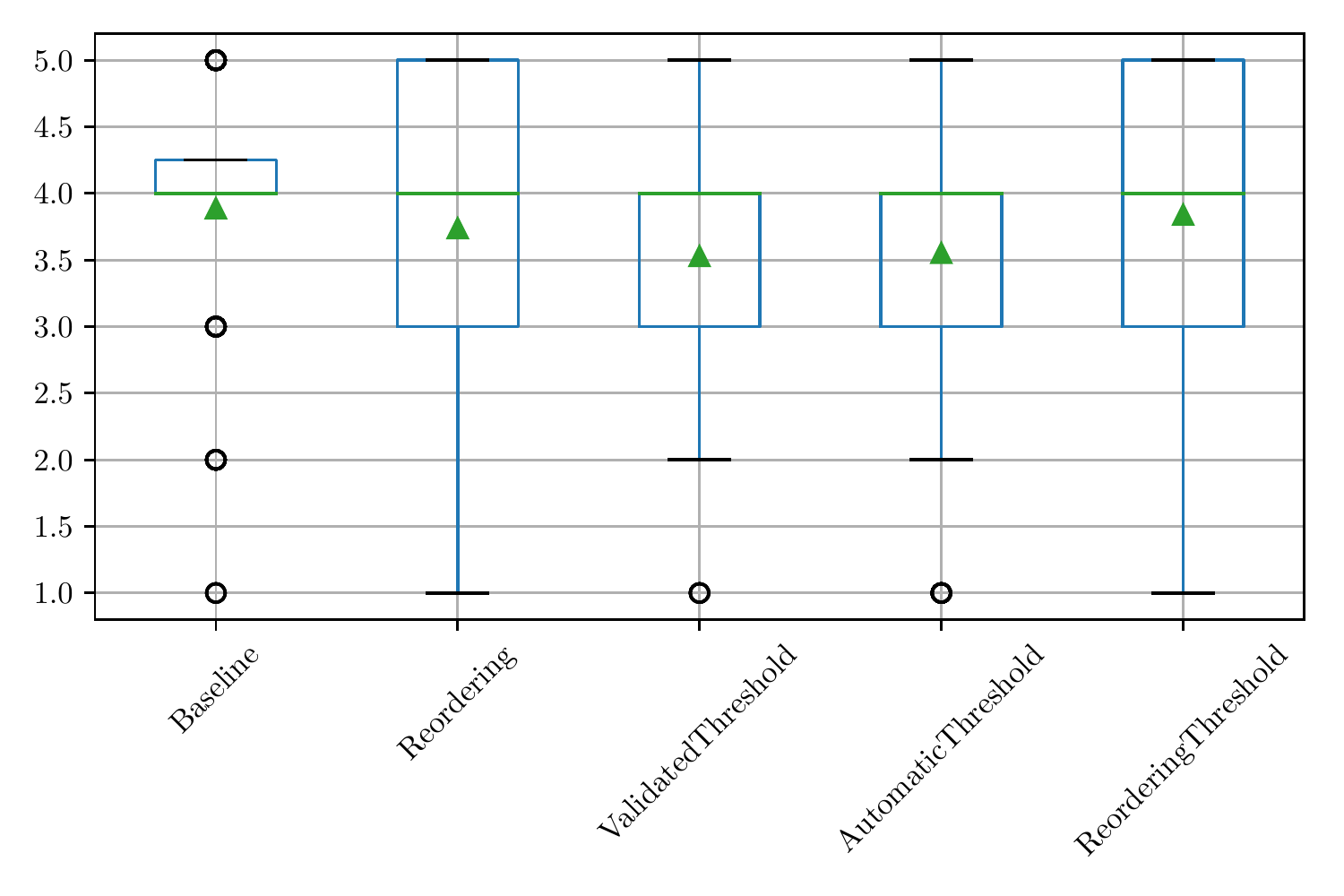}
		\caption{User Beliefs}
		\label{fig:bpg2}
	\end{subfigure}\hfil
	\begin{subfigure}{0.25\textwidth}
		\includegraphics[width=\linewidth]{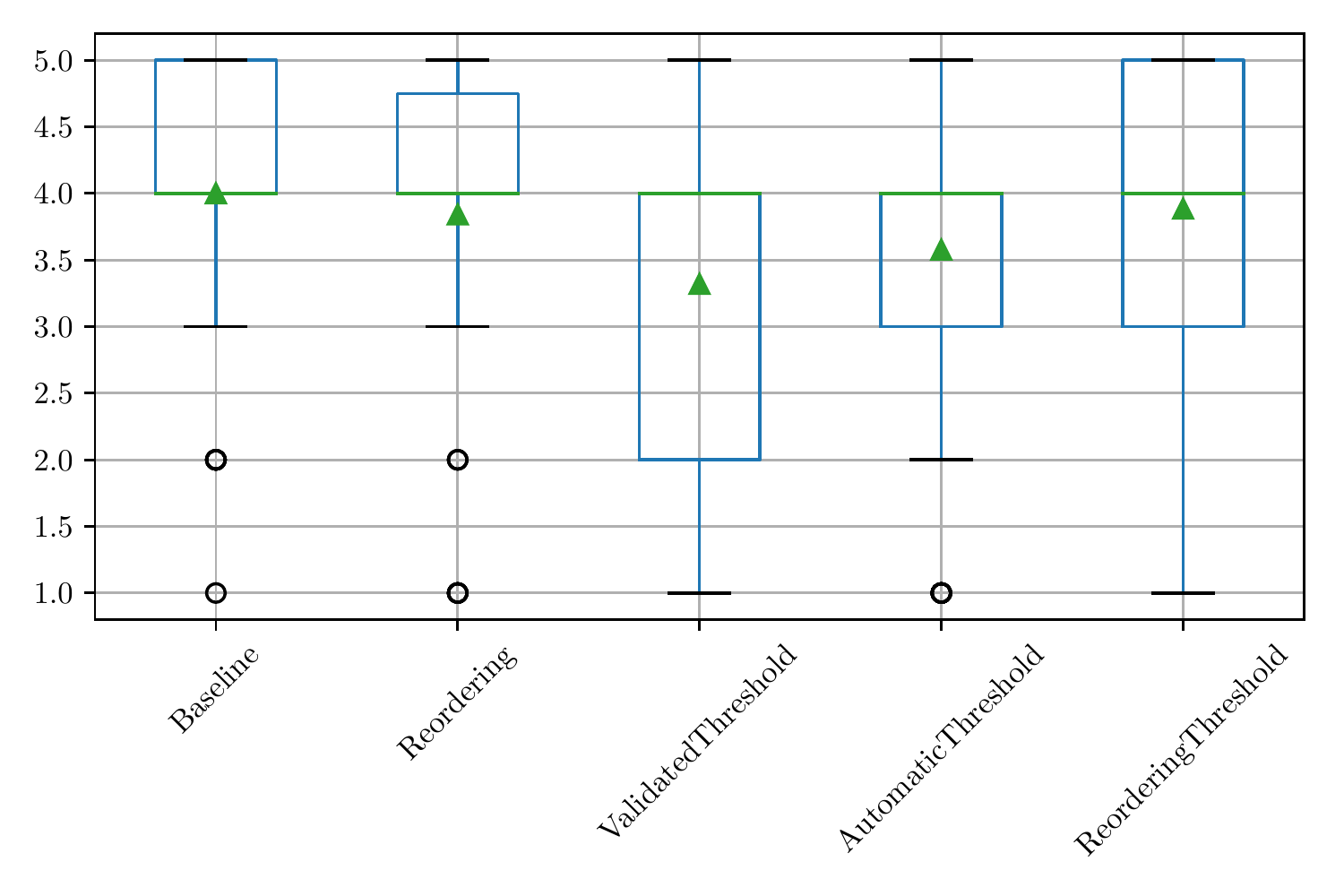}
		\caption{User Attitudes}
		\label{fig:bpg3}
	\end{subfigure}
	\caption{Grouped Boxplots for the Adapted ResQue Questionnaire}
	\label{fig:resque-boxplot-grouped}
\end{figure}

The boxplotes shown in Figure~\ref{fig:resque-boxplot-grouped} are generated for the grouped question topics (cf. Table~\ref{tab:ResQueGroups}).

\end{document}